\documentclass[epj]{svjour}
\usepackage{graphics}
\usepackage{mathrsfs}  
\usepackage{adjustbox}
\usepackage[all]{xy}
\usepackage{epsfig, cancel}
\usepackage{cite}
\usepackage{latexsym}
\usepackage{url}
\usepackage{dcolumn}
\usepackage{color}
\usepackage{amsfonts,amssymb,amsmath, txfonts}
\usepackage{graphicx,epsfig}
\usepackage{psfrag}
\usepackage{subfigure}
\usepackage{hyperref}
\hypersetup{colorlinks=true}
\usepackage{mathtools}
\usepackage{enumitem}
\usepackage{float}
\usepackage[dvipsnames]{xcolor}
\usepackage{xcolor}
\hypersetup{ linktoc=all,
    colorlinks, linkcolor={brightpink},
    citecolor={blue}, urlcolor={blue}
}
\definecolor{rosy}{RGB}{230,235,252}
\definecolor{myframetitle}{RGB}{90,89,170}
\definecolor{myblocktitle}{RGB}{140,185,249}
\definecolor{mytitle}{RGB}{10,80,26}

\definecolor{darkgreen}{RGB}{27,130,45}
\definecolor{darkblue}{rgb}{0,0,0.3}
\definecolor{darkred}{rgb}{0.7,0,0}

\definecolor{light gray}{RGB}{220,220,220}
\definecolor{dark purple}{RGB}{108,0,217}
\definecolor{pink}{RGB}{190,20,100}
\definecolor{orang}{RGB}{193,63,0}
\definecolor{green}{RGB}{11,98,17}
\definecolor{darkpink}{RGB}{153,0,76}
\definecolor{bluegreen}{RGB}{0,102,102}
\definecolor{greenlagan}{RGB}{0,102,0}
\definecolor{redgreen}{RGB}{102,102,0}
\definecolor{Redgreen}{RGB}{153,76,0}
\definecolor{vividviolet}{rgb}{0.62, 0.0, 1.0}
\definecolor{amaranth}{rgb}{0.9, 0.17, 0.31}
\definecolor{palatinateblue}{rgb}{0.15, 0.23, 0.89}
\definecolor{brightpink}{rgb}{1.0, 0.0, 0.5}
\definecolor{cornflowerblue}{rgb}{0.39, 0.58, 0.93}
\definecolor{deepcarminepink}{rgb}{0.94, 0.19, 0.22}
\definecolor{radicalred}{rgb}{1.0, 0.21, 0.37}

\def\H0{{\text{H}\hspace*{-2.05mm}\text{H} 0\hspace*{-1.35mm}0\ }}

\def\be{\begin{equation}}
\def\ee{\end{equation}}
\def\beq{\begin{equation}}
\def\eeq{\end{equation}}
\def\bea{\begin{eqnarray}}
\def\eea{\end{eqnarray}}

\begin{document}
\title{A comparison of Bayesian and frequentist confidence intervals in the presence of a late Universe degeneracy}

\author{Eoin \'O Colg\'ain\inst{1} \and Saeed Pourojaghi \inst{2} \and M. M. Sheikh-Jabbari \inst{2, 3} \and Darragh Sherwin\inst{1}
}                     
%
\institute{Atlantic Technological University, Ash Lane, Sligo, Ireland \and 
School of Physics, Institute for Research in Fundamental Sciences (IPM), P.O.Box 19395-5531, Tehran, Iran \and The Abdus Salam ICTP, Strada Costiera 11, I-34014 Trieste, Italy}
\date{Received: date / Revised version: date}
%
\abstract{
Hubble tension is a problem in one-dimensional (1D) posteriors, since local $H_0$ determinations are only sensitive to a single parameter. Projected 1D posteriors for $\Lambda$CDM cosmological parameters become more non-Gaussian with increasing effective redshift when the model is fitted to redshift-binned data in the late Universe. We explain mathematically why this non-Gaussianity arises and show using observational Hubble data (OHD) that Markov Chain Monte Carlo (MCMC) marginalisation leads to 1D posteriors that fail to track the $\chi^2$ minimum at $68\%$ confidence level in high redshift bins. {To gain a second perspective, }
we resort to profile {likelihoods} as a complementary technique. Doing so, we observe that $z \gtrsim 1$ cosmic chronometer (CC) data currently prefers a non-evolving (constant) Hubble parameter over a Planck-$\Lambda$CDM cosmology at $\sim 2 \sigma$. Within the Hubble tension debate, it is imperative that subsamples of data sets with differing redshifts yield similar $H_0$ values. In addition, we confirm that MCMC degeneracies observed in 2D posteriors are not due to curves of constant $\chi^2$. Finally, on the assumption that the Planck-$\Lambda$CDM cosmological model is correct, using profile {likelihoods} we confirm  a $>2 \sigma$ discrepancy with Planck-$\Lambda$CDM in a combination of  CC and baryon acoustic oscillations (BAO) data beyond $ z \sim 1.5$. This confirms a discrepancy reported earlier with fresh methodology.
} 
\maketitle

\section{Introduction}
\label{sec:intro}

The flat $\Lambda$CDM model is the minimal model that fits Cosmic Microwave Background (CMB) data. Remarkably, CMB data from the Planck satellite \cite{Planck:2018vyg} constrains the $\Lambda$CDM model to sub-percent errors, thereby not only providing the strongest constraints, but also a prediction for cosmological probes in the late Universe. The unmitigated success of the $\Lambda$CDM model is that CMB, Type Ia supernovae (SNe) \cite{Riess:1998cb, Perlmutter:1998np} and baryon acoustic oscillations (BAO) \cite{Eisenstein:2005su} agree on a $\Lambda$CDM Universe that is approximately $30 \%$ pressureless matter. Thus, one key prediction of the Planck-$\Lambda$CDM model agrees across early and late Universe cosmological probes. Given this non-trivial agreement, any discrepancies that arise elsewhere constitute puzzles. 

Nevertheless, one cannot define any \textit{model} for a dynamical system, especially a complicated system like the Universe, using data from a cosmic snapshot.\footnote{Here, we mean CMB data with an effective redshift $z \sim 1100$.} At best, one has a \textit{prediction} and not a model. In recent years, key predictions of Planck data have been challenged by late Universe determinations of the Hubble constant $H_0$ \cite{Riess:2021jrx, Freedman:2021ahq, Pesce:2020xfe, Blakeslee:2021rqi, Kourkchi:2020iyz} and the $S_8:= \sigma_8 \sqrt{\Omega_m/0.3}$ parameter \cite{HSC:2018mrq, KiDS:2020suj, DES:2021wwk, Boruah:2019icj, Said:2020epb}. Given the diversity of the late Universe probes (see reviews \cite{Perivolaropoulos:2021jda, Abdalla:2022yfr}), it is highly unlikely that any single systematic can be found to explain the discrepancies. That being said, in astrophysics one can never preclude systematics; 3 decades after Phillips' seminal paper \cite{Phillips:1993ng}, we are still debating an ad hoc correction for the mass of the host galaxy in Type Ia SNe \cite{NearbySupernovaFactory:2018qkd, Kang:2019azh, Brout:2020msh, Lee:2021txi}. Bearing in mind that Type Ia SNe are one of our best understood cosmological probes, one quickly understands that any systematics debate may be endless. 

Thus, it may be far more expedient to assume that the $\Lambda$CDM model is breaking down and to look for tell-tale signatures of model breakdown. If signatures cannot be found, one arrives at a contradiction, and revisits the assumption that the model is breaking down. For physicists, \textit{model breakdown comes about when model fitting parameters return discrepant values at different time slices or epochs}. Translated into astronomy, this equates to discrepant cosmological parameters in different redshift ranges. The usual $H_0, S_8$ tensions  may also be viewed in the same light; a discrepancy between high and low redshift inferences/measurements of the parameters \cite{Perivolaropoulos:2021jda, Abdalla:2022yfr}. Nevertheless, early and late Universe observables are typically not the same, so one is confronted with a rich set of potential systematics. 

Within the context of the $\Lambda$CDM model, the Hubble constant $H_0$ arises \textit{mathematically} as an integration constant when one solves the Friedmann equations. This provides a mathematical perspective on $H_0$ \cite{Krishnan:2020vaf, Krishnan:2022fzz}. Nevertheless, there is a second observational perspective; $H_0$ is a fitting parameter in the $\Lambda$CDM model. As a result, one can perform a consistency check of the $\Lambda$CDM model by confronting it to observational data in different redshift ranges or cosmological epochs \cite{Krishnan:2020vaf, Krishnan:2022fzz, Colgain:2022nlb,Colgain:2022rxy}. If one recovers the same $H_0$ value at all redshifts, then mathematics and observation are in sync. In contrast, if one does not, the data and the $\Lambda$CDM model are no longer in agreement. Similarly, $\rho_{m0}=H_0^2\Omega_m$, an integration constant of the matter continuity equation, implies matter density $\Omega_m$ is a mathematically constant quantity, so the same logic applies. In the late Universe within the $\Lambda$CDM model, $H_0$ is correlated with matter density $\Omega_m$, 
while $\Omega_m$ is correlated with $S_8 \propto \sigma_8 \sqrt{\Omega_m}$. Thus, there is at least one simple scenario, namely redshift evolution of cosmological fitting parameters in the late Universe, where ``$H_0$ tension'' and ``$S_8$ tension'' are not independent and simply symptoms of $\Lambda$CDM model breakdown \cite{Colgain:2022nlb, Colgain:2022rxy, Colgain:2022tql}. 

The next relevant question is, where is the evidence for evolving cosmological fitting parameters in the late Universe? Starting with strong lensing time delay \cite{Wong:2019kwg, Millon:2019slk},\footnote{Systematics are explored in \cite{Millon:2019slk} and the descending trend is not an obvious systematic. The lensed system RXJ1131-1231 \cite{Sluse:2003iy}, which partly drives the trend, has recently been re-analysed using spatially resolved stellar kinematics of the host galaxy \cite{Shajib:2023uig}, and the higher $H_0$ value remains robust, admittedly with inflated errors. As TDCOSMO project to analyse 40 lenses, the prospect of a discovery of a descending $H_0$ trend assuming the $\Lambda$CDM model remain strong.} descending trends of $H_0$ with redshift have been reported in Type Ia SNe \cite{Dainotti:2021pqg, Colgain:2022nlb, Colgain:2022rxy,  Malekjani:2023dky, Hu:2022kes, Jia:2022ycc} and combinations of data sets \cite{Krishnan:2020obg, Dainotti:2022bzg}. On the other hand, larger values of $\Omega_m$ have been noted in high redshift observables, primarily quasars (QSOs) \cite{Risaliti:2015zla, Risaliti:2018reu, Lusso:2020pdb, Yang:2019vgk, Khadka:2020vlh, Khadka:2020tlm, Khadka:2021xcc, Pourojaghi:2022zrh},\footnote{Just as with Type Ia SNe, the systematics of QSOs are being investigated \cite{Zajacek:2023qjm}.} but also Type Ia SNe \cite{Colgain:2022nlb, Colgain:2022rxy, Malekjani:2023dky, Pasten:2023rpc} (see also \cite{Wagner:2022etu, Sakr:2023hrl}). Note, as emphasised earlier, if $H_0$ as a fitting parameter evolves at the background level, which is only possible if the $\Lambda$CDM model has broken down, correlated fitting parameters are expected to also evolve. Moreover, mock analysis within the $\Lambda$CDM setting reveals that evolution of best fit $(H_0, \Omega_m)$ parameters cannot be precluded, and conversely possesses a finite likelihood, in either observational Hubble data (OHD) \textit{or} angular diameter distance data \textit{or} luminosity distance data \cite{Colgain:2022tql}. We stress that this result \textit{rests on mock analysis}; it represents a purely mathematical statement about the $\Lambda$CDM model that is independent of systematics. 

Separately, at the perturbative level, redshift evolution of $S_8$ or $\sigma_8$ has been reported in galaxy cluster number counts and Lyman-$\alpha$ spectra \cite{Esposito:2022plo}, $f \sigma_8$ constraints from peculiar velocities and redshift space distortions (RSD) \cite{Adil:2023jtu}, comparison between weak \cite{HSC:2018mrq, KiDS:2020suj, DES:2021wwk} and CMB lensing \cite{ACT:2023dou, ACT:2023kun} (see also \cite{Tutusaus:2023aux}). What is important here is that these observations appear to restrict the evolution in $S_8$ to the late Universe. In \cite{ACT:2023ipp} the possibility was raised that \textit{``tracers at higher redshift and probing larger scales prefer higher $S_8$''}.\footnote{There are also conflicting observations of high redshift $\sigma_8$ or $S_8$ values that are lower than Planck in the late Universe \cite{Miyatake:2021qjr, Alonso:2023guh}, so either this trend is not universal, or systematics are at play.} Nevertheless, one can argue against evolution with scale on the grounds that cosmic shear \cite{HSC:2018mrq, KiDS:2020suj, DES:2021wwk}, which is sensitive to smaller scales (larger $k$), and peculiar velocity constraints \cite{Boruah:2019icj, Said:2020epb}, which are sensitive to larger scales (smaller $k$), both prefer lower values of $S_8$. Moreover, both galaxy clusters and Lyman-$\alpha$ spectra are expected to probe similar scales.\footnote{We thank Matteo Viel for correspondence on this point.} Thus, if systematics are not impacting results, then redshift evolution is the only point of agreement in the observations \cite{Esposito:2022plo, Adil:2023jtu, HSC:2018mrq, KiDS:2020suj, DES:2021wwk, ACT:2023dou, ACT:2023kun, ACT:2023ipp}. Note also that redshift is more fundamental than scale in FLRW cosmology; one must solve the Friedmann equations in either time or redshift before one contemplates any discussion of scale.  

The purpose of this letter is to revisit the analysis presented in \cite{Colgain:2022rxy,Colgain:2022tql}, where the evidence for evolution in fitting parameters was quantified on the basis of mock simulations and not Markov Chain Monte Carlo (MCMC), the technique most familiar in cosmology. The fundamental problem is that once one bins low redshift data and studies evolution of cosmological fitting parameters with bin redshift, one quickly encounters degeneracies and projection effects in MCMC analyses. These effects are not just the preserve of exotic models \cite{Herold:2021ksg, Gomez-Valent:2022hkb, Meiers:2023gft}, such as Early Dark Energy (EDE) \cite{Poulin:2018cxd, Niedermann:2019olb}, and happen in the simplest model when one bins data. The most striking demonstration of the {effect} is that the peaks of MCMC posteriors no longer coincide with the minimum of the likelihood (see \cite{Gomez-Valent:2022hkb}). Ultimately,  {this outcome may be} expected  because one is working in a regime of the $\Lambda$CDM model with projected 1D non-Gaussian probability distributions   \cite{Colgain:2022tql}  (see also \cite{Colgain:2022rxy}). Note, this is not a problem with MCMC marginalisation \textit{per se}, but a deeper problem that impacts MCMC marginalisation. We remind the reader once again that Hubble tension is a problem in 1D posteriors, so projection cannot be avoided if the goal is to compare cosmological and local $H_0$ constraints.

 The structure of this paper is as follows. In section \ref{sec:MCMC_bias} we {identify the problem prompted by a late Universe degeneracy in the $\Lambda$CDM model}. In section \ref{sec:PD} we introduce profile distributions (PDs) \cite{Gomez-Valent:2022hkb}, {a mild variant of frequentist profile likelihoods \cite{Trotta:2017wnx},} as a means of addressing the {effect} and confirm that the statistical significance of discrepancies from mock simulations agree well with PD analysis. In section \ref{sec:MCMC_puzzle} we study whether a degeneracy in MCMC posteriors in the Bayesian approach translates into a constant $\chi^2$ curve in the frequentist approach. In section \ref{sec:tension}, we revisit and confirm the high redshift OHD tensions reported in \cite{Colgain:2022rxy}. We end in section \ref{sec:discussion} with concluding remarks. 

\section{{A late Universe degeneracy}}
\label{sec:MCMC_bias}
In this section we illustrate a problem that arises in the (flat) $\Lambda$CDM model when OHD is binned by redshift. Our treatment through to the end of section \ref{sec:CCbias} overlaps with \cite{Colgain:2022tql}, where exclusively mock data was considered. Readers familiar with the observations in \cite{Colgain:2022tql} may skip directly to subsection \ref{sec:features}. Subsection \ref{sec:features} onwards discusses features in observed cosmic chronometer (CC) data. Our analysis in subsection \ref{sec:features} onwards overlaps with \cite{Colgain:2022rxy} but is distinct due to the absence of external priors and BAO data. We reintroduce priors/BAO later in section \ref{sec:tension} to comment on a discrepancy with the Planck-$\Lambda$CDM cosmology and how the results compare with \cite{Colgain:2022rxy}. 

Returning to our main focus, the basic problem is that as one fits $H(z)$ constraints in bins of increasing effective redshift, the late Universe $\Lambda$CDM Hubble parameter $H(z)$ transitions from a model with two degrees of freedom $(H_0, \Omega_m)$ to a model where only a single degree of freedom $\Omega_m h^2$ ($h:= H_0/100$) is well constrained. In the cosmology literature, this is presented as the idea that ``\textit{high redshift data can only constrain} $\Omega_m h^2$". This statement is supported by banana-shaped contours in $(H_0, \Omega_m)$ MCMC posteriors. Throughout, little consideration has been given to whether Bayesian and frequentist methods agree on the ``high redshifts" where the late Universe $\Lambda$CDM model loses the additional degree of freedom. Noting that the transition is inevitable, but gradual, the point of this paper is to explore whether Bayesian and frequentist methods lose sensitivity to the full parameter space at the same rate or in step. We will demonstrate that this is not the case and that frequentist methods can identify constraints in redshift ranges where MCMC posteriors are unconstrained and thus inconclusive. Alternatively put, within our assumptions and setup, we demonstrate the frequentist methods lead to better constrained confidence intervals than Bayesian credible intervals when a degeneracy is present. Moreover, unconstrained MCMC posteriors manifest themselves as projection effects in marginalised 1D posteriors, which biases MCMC inferences. Separately, we confirm that in settings where the 2D parameter space is well constrained, as is evident by Gaussian MCMC posteriors, that Bayesian and frequentist methods agree.

\subsection{Mathematical Arguments}
\label{sec:math}
Consider an exercise where one bins OHD and confronts it to the $\Lambda$CDM Hubble Parameter $H(z)$ in the late Universe, a setting where the radiation sector can be safely decoupled. In high redshift bins ($z \gg 0$) in the matter-dominated regime, the Hubble parameter becomes insensitive to the dark energy (DE) sector: 
\be
\label{eq:lcdm}
H(z) = H_0 \sqrt{1-\Omega_m + \Omega_m (1+z)^3} \xrightarrow[z \gg 0]{} H_0 \sqrt{\Omega_m} (1+z)^{\frac{3}{2}}.  
\ee
More concretely, taking a large $z$ limit we see that high redshift data can only constrain the combination $\rho_{m0}=H_0^2{\Omega_m}$. For data in a redshift bin with large $z$, $z \gg 0$, this means that one can only constrain the combination $\Omega_m h^2$, but $H_0$ and $\Omega_m$ remain largely unconstrained. Alternatively put, for any given $\Omega_m h^2$ constraint, there is {a large number of permissible} $(H_0, \Omega_m)$ pairs. Translated into a probability density function (PDF), this is simply the statement that in a very high redshift bin {with $z \gg 0$}, one expects {almost} uniform/flat distributions for $H_0$ and $\Omega_m$ with the model (\ref{eq:lcdm}).  

Here what is important to us is the observation that these PDFs must flatten in a non-Gaussian manner. To appreciate this fact, we observe that high redshift OHD only constrains $\Omega_m h^2$ well.\footnote{Note that observables like SNe or QSOs that measure $D_L(z)=c (1+z)\int_0^z \textrm{d} z'/H(z')$ are mainly sensitive to the low redshift part of $H(z)$, i. e. the combination $H_0^2 (1-\Omega_m)$, and in this sense they are complementary to the OHD data which is more sensitive to high redshift part of $H(z)$, $H_0^2\Omega_m$. The complementarity can be demonstrated by combining $H(z)$ and $D_{L}(z)$ constraints and checking that one recovers mock data input parameters in all redshift bins \cite{Colgain:2022tql}. } For this reason, best fit parameters are constrained to a $\Omega_m h^2 = \textrm{constant}$ curve in the $(H_0, \Omega_m)$-plane. The almost flat $H_0$ and $\Omega_m$ PDFs can only arise if this curve stretches in the $(H_0, \Omega_m)$-plane. As a result of this stretching, one ends up with a relatively uniform distribution on a curve. At the extremes of the curve, one finds a distribution of large $H_0$ values, which do not differ greatly in $\Omega_m$, and they get projected to a peak at small values on the $\Omega_m$ axis. Conversely, at the other end of the curve, one finds a distribution of small $\Omega_m$ values, which do not differ greatly in $H_0$, and they get projected onto a peak at large values on the $H_0$ axis.  This is a ``projection effect'' in common cosmology parlance.  It is driven by the irrelevance of the DE sector at high redshift and the constraint $\Omega_m h^2 = \textrm{constant}$ from the $\Lambda$CDM model (\ref{eq:lcdm}). Together these features distort {projected 1D distributions} away from a Gaussian configuration. 

Thus, simply by binning and fitting OHD to the $\Lambda$CDM model one enters a non-Gaussian regime as the effective redshift of the bin increases. {As explained, this non-Gaussianity can be traced to the $\Lambda$CDM model transitioning from a 2D model at lower redshifts to an effective 1D model at higher redshifts.} This effect, which is expected from the purely mathematical arguments above, has been confirmed in mock data \cite{Colgain:2022rxy, Colgain:2022tql}, and in line with expectations, we demonstrate how it impacts MCMC inferences with observed data in the next subsection.  

\begin{figure}[htb]
   \centering
\includegraphics[width=80mm]{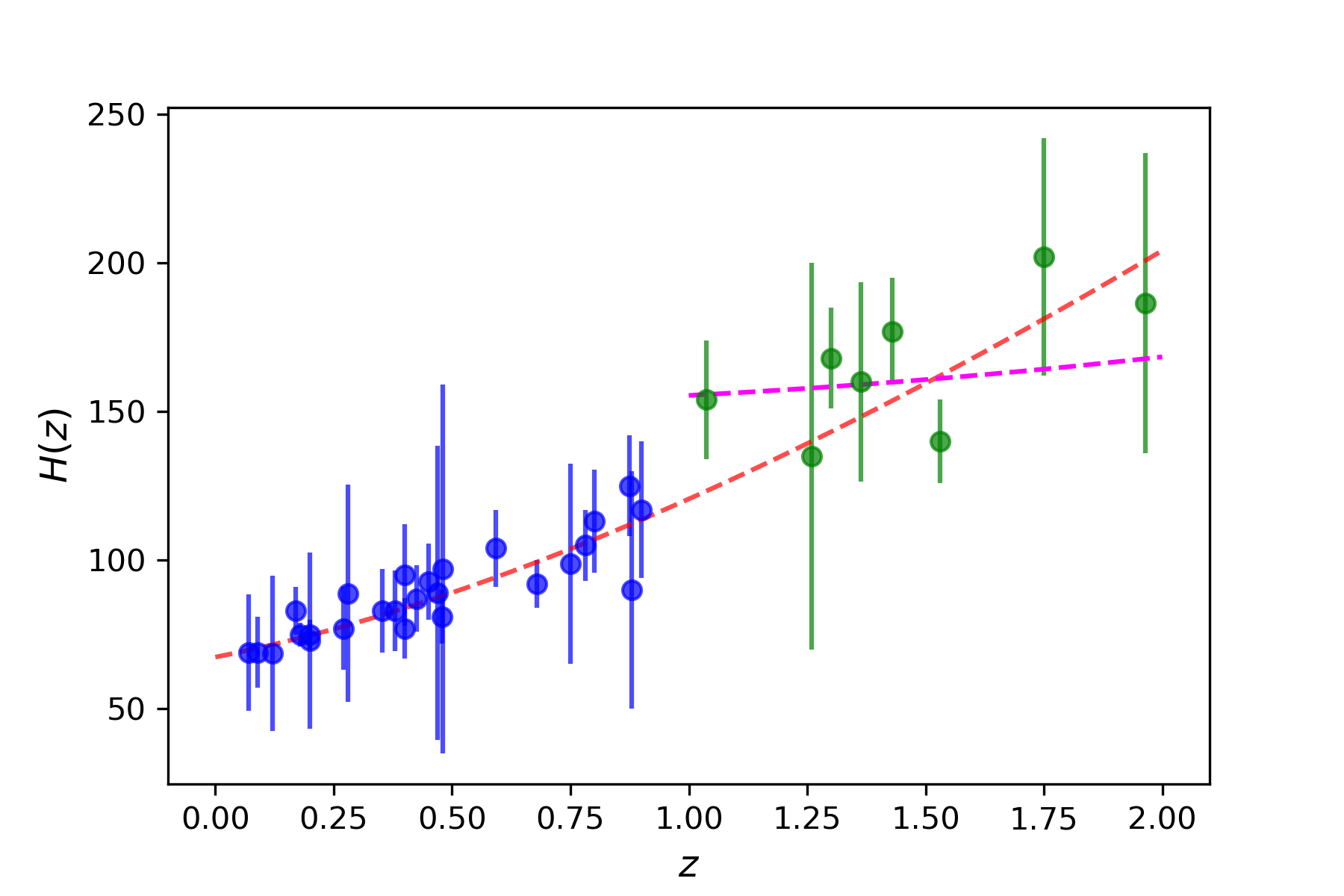}
\caption{The CC data split above and below $z=1$ alongside the Planck-$\Lambda$CDM $H(z)$ (dashed red) and best fit $H(z)$ (dashed magenta) above $z=1$. From the high redshift data (green), it is evident that a horizontal $H(z)$ corresponding to $\Omega_m \sim 0$ is preferred.}
\label{fig:CC} 
\end{figure}

\subsection{Cosmic Chronometer (CC) Data}
\label{sec:CCbias}
Here we work with OHD from the cosmic chronometer (CC) program \cite{Jimenez:2001gg}. Concretely, we study 34 $H(z)$ constraints spanning the redshift range $0.07 \leq z \leq 1.965$ \cite{Stern:2009ep, Moresco:2012jh, Zhang:2012mp, Moresco:2016mzx, Ratsimbazafy:2017vga, Borghi:2021rft, Jiao:2022aep, Tomasetti:2023kek}. We illustrate the data in Fig.~\ref{fig:CC}, where it is consistent with Fig. 9 of \cite{Tomasetti:2023kek} modulo the fact that we have an additional data point at $z = 0.8$, which is not independent. See Table 1.1 of \cite{Moresco:2023zys} for a discussion on the independence of the data points. While CC data may eventually be good enough to arbitrate on Hubble tension \cite{Moresco:2023zys}, the data is not good enough on its own to do cosmology. To put this comment in context, we observe that the errors in Fig.~\ref{fig:CC} do not include systematic errors (see \cite{Moresco:2020fbm} for an account of the systematics). As a result the constraints we get on cosmological parameters will be underestimated. Thus, from our perspective the data in Fig.~\ref{fig:CC} is simply some representative cosmological data in the OHD class, providing a concrete setting where we can compare Bayesian and frequentist methods.

\paragraph{Methodology:} We impose a low redshift cut-off on the OHD $z_{\textrm{min}}$, removing all data points with redshifts $z_i < z_{\textrm{min}}$, and then extremise the $\chi^2$ likelihood, 
\be
\label{eq:chi2}
\chi^2 = Q^{T} \cdot C^{-1} \cdot Q, 
\ee
where $C$ is the covariance matrix, which is simply the square of the $H_i$ errors on the diagonal, and $Q$ is the vector, 
\be
\label{eq:Q}
Q_i = H_i - H_{\textrm{model}}(z_i), 
\ee
where $H_i:=H(z_i)$ denotes OHD and $H_{\textrm{model}}(z)$ is the model (\ref{eq:lcdm}) without the high redshift limit. The best fit $(H_0, \Omega_m)$ parameters correspond to the minimum of the $\chi^2$, while on the assumption of Gaussian errors, we estimate the errors from a Fisher matrix (appendix \ref{sec:fisher}). In parallel, we perform MCMC marginalisation through \textit{emcee} \cite{Foreman-Mackey:2012any}. More concretely, subject to the priors $H_0 \in [0, 200 ]$ and $\Omega_m \in [ 0, 1]$, the latter restricting us to a physical regime, we record {the peak and $68\%$ credible interval from projected 1D $H_0$ and $\Omega_m$ posteriors. \footnote{We define the $68\%$ credible interval by converting the 1D posterior into a histogram. Therein, each bin centered on a given $H_0$ or $\Omega_m$ value has a width $w_i$ and a count $c_i$. The total area of the \textit{discretised} 1D posterior (histogram) is simply $A_{\textrm{tot}} = \sum_{i} w_i c_i$, where the sum ranges over all bins. Starting from the maximum count (posterior peak) in the histogram $c_{\textrm{max}}$, we identify the subset of bins with counts $c_j$ greater than a fixed value so that the normalised area is 68\% of the total, i. e. $\sum_{j} w_j c_j/A_{\textrm{tot}} \sim 0.68$. The range of $H_0$ or $\Omega_m$ values covered by this subset of bins then defines the $68 \%$ credible interval. We have checked that our methodology agrees with \textit{GetDist} \cite{Lewis:2019xzd} and recovers the standard practice of quoting $16^{\textrm{th}}$, $50^{\textrm{th}}$ (peak) and $84^{\textrm{th}}$ percentiles for Gaussian posteriors.} We stress that our Fisher matrix analysis assumes that 1D posteriors are Gaussian. As we shall see from our MCMC analysis, this is a good approximation when $z_{\textrm{min}} = 0$, but as $z_{\textrm{min}}$ increases, 1D posteriors becomes more non-Gaussian due to a degeneracy giving rise to a banana-shaped contour in the 2D $(H_0, \Omega_m)$ posterior. The na\"ive interpretation of this result is that the data is not good enough to constrain the model. As we shall argue, this interpretation {may only be true in the Bayesian framework}. Nevertheless, by evaluating the MCMC chains on the $\chi^2$ likelihood (\ref{eq:chi2}), we confirm that the parameters corresponding to the minimum $\chi^2$ are tracking the best fit from the Fisher matrix analysis. This provides a non-trivial consistency check between the Fisher matrix analysis based on optimisation (gradient descent) and MCMC marginalisation, where the $\chi^2$ is simply evaluated on the MCMC chain.}

\begin{table*}
    \centering
    \begin{tabular}{|c|c|c|c|c|c|}
\hline
    \rule{0pt}{3ex} $z_{\textrm{min}}$ & \# CC & \multicolumn{2}{|c|}{Fisher Matrix}  & \multicolumn{2}{|c|}{MCMC} \\
    \hline
    \rule{0pt}{3ex} & & $H_0$ (km/s/Mpc) & $\Omega_m$ & $H_0$ (km/s/Mpc) & $\Omega_m$ \\
    \hline\hline
    \rule{0pt}{3ex} $0$ & $34$ & $68.14 \pm 3.07$ & $0.320 \pm 0.059$ & ${68.01^{+2.77}_{-3.33}}$  ($68.12$) & ${0.315^{+0.072}_{-0.045}}$ ($0.321$) \\
    \hline 
    \rule{0pt}{3ex} $0.2$ & $27$ & $65.03 \pm 6.65$ & $0.368 \pm 0.118$ & ${63.98^{+6.16}_{-7.82}}$ ($64.98$) & ${0.364^{+0.150}_{-0.110}}$ ($0.369$) \\
    \hline 
    \rule{0pt}{3ex} $0.4$ & $22$ & $62.42 \pm 8.38$ & $0.411 \pm 0.161$ & ${59.87^{+7.82}_{-9.65}}$ ($62.39$) & ${0.400^{+0.212}_{-0.143}}$ ($0.411$)\\
    \hline 
     \rule{0pt}{3ex} $0.6$ & $15$ & $59.83 \pm 17.21$ & $0.454 \pm 0.338$ & ${46.79^{+21.00}_{-3.05}}$ ($59.86$) & ${0.482^{+0.370}_{-0.248}}$ ($0.453$) \\
    \hline 
     \rule{0pt}{3ex} $0.7$ & $14$ & $79.11 \pm 19.40$ & $0.222 \pm 0.162$ & ${59.22^{+22.31}_{-14.07}}$ ($79.18$) & ${0.181^{+0.319}_{-0.095}}$ ($0.222$) \\
    \hline 
      \rule{0pt}{3ex} $0.8$ & $11$ & $103.97 \pm 24.94$ & $0.097 \pm 0.088$ & ${83.13^{+25.15}_{-37.72}}$ ($104.02$) & ${0.088^{+0.285}_{-0.070}}$ ($0.096$) \\
    \hline 
     \rule{0pt}{3ex} $1$ & $8$ & $150.37 \pm 31.21$ & $0.010 \pm 0.035$ & ${120.54^{+33.14}_{-75.33}}$ ($150.38$) & ${<0.175}$ ($0.010$) \\
    \hline 
     \rule{0pt}{3ex} $1.2$ & $7$ & $154.35 \pm 42.95$ & $0.006 \pm 0.042$ & ${45.57^{+97.02}_{-4.62}}$ ($154.47$) & ${<0.390}$ ($0.006$) \\
    \hline 
    \rule{0pt}{3ex} $1.4$ & $4$ & $125.41 \pm 79.55$ & $0.039 \pm 0.132$ & ${45.08^{+42.54}_{-6.96}}$ ($125.44$) & ${< 0.530}$ ($0.039$) \\
    \hline 
    \rule{0pt}{3ex} $1.5$ & $3$ & $36.12 \pm 72.69$ & $1.000 \pm 4.269$ & ${42.05^{+32.35}_{-7.70}}$ ($36.16$) & ${<0.570}$ ($0.999$)\\ \hline
    \end{tabular}
    \caption{{Comparison between Fisher matrix and MCMC analysis for CC data with a low redshift cut-off $z_{\textrm{min}}$. We record the number of data points, the minimum of the $\chi^2$ (Fisher Matrix), the peak of the posteriors (MCMC), the $68\%$ ($1 \sigma$) intervals estimated from both the Fisher matrix and MCMC posteriors, and the minimum $\chi^2$ from the MCMC chain in brackets. MCMC marginalisation exhibits non-Gaussian $68\%$ credible intervals, and for $z_{\textrm{min}} > 1$, the minimum value of the $\chi^2$ from the MCMC chain falls outside of this interval. The $\chi^2$ minimum from the MCMC chain agrees with direct $\chi^2$ minimisation up to small numbers in line with expectations.} }
    \label{tab:LCDM_CC}
\end{table*}

\subsection{Features in CC Data}
\label{sec:features}
From Table \ref{tab:LCDM_CC} we see that there are noticeable trends in CC data when it is binned. Beginning with the best fits corresponding to the minimum of the $\chi^2$, from $z_{\textrm{min}} = 0$ through to $z_{\textrm{min}} = 0.6$, there is a decreasing trend in best fit $H_0$ values, which is compensated by an increasing trend in best fit $\Omega_m$ values. From Fig.~\ref{fig:CC} it is difficult to visibly discern any trend from the raw data. From $z_{\textrm{min}} = 0.7$ through to $z_{\textrm{min}} = 1.4$, there is in contrast a preference for larger best fit $H_0$ and smaller best fit $\Omega_m$ values. This trend is evident from the raw data, where at higher redshifts one sees large scatter and large fractional errors in the data. For $z_{\textrm{min}} = 1$, it is clear that the best fit line in magenta corresponding to $(H_0, \Omega_m) = (150.4, 0.01)$ ($z_{\textrm{min}} = 1$ entry in Table \ref{tab:LCDM_CC}) is closer to a horizontal line than the Planck-$\Lambda$CDM cosmology in red. \footnote{The large $H_0$ value is compensated by a low $\Omega_m$ leading to an age of the  Universe that is consistent with Planck.} To be more explicit, for $z_{\textrm{min}} = 0$, $\rho_{m0}:=H_0^2\Omega_m\simeq 1500$ which is close to the Planck value, whereas for $z_{\textrm{min}} = 1$, $\rho_{m0}\simeq 225$. The sharp drop in $\rho_{m0}$ means the magenta line should be almost horizontal. Finally, for $z_{\textrm{min}} = 1.5$, we switch to an opposite regime of parameter space with unexpectedly low and high {best fit} values of $H_0$ and $\Omega_m$, respectively, a trend which is evident in the data, but there are only three data points. Despite, the small number of data points, the tendency for smaller $H_0$ and larger $\Omega_m$ inferences within $\Lambda$CDM cosmology at high redshifts has been documented across three independent observables \cite{Colgain:2022rxy}. We will come back to this claim in section \ref{sec:tension}. 

\begin{figure}[htb]
   \centering
\includegraphics[width=80mm]{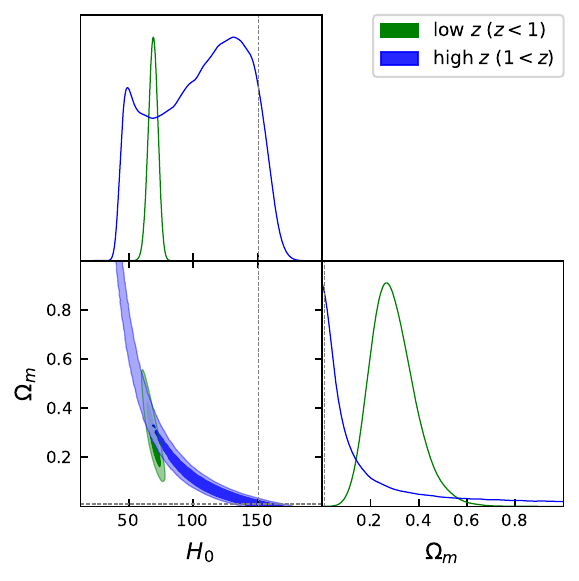}
\caption{Cosmological parameter inferences from MCMC chains for the CC data split at $z=1$. Dashed lines denote the minimum of the $\chi^2$ for $z > 1$ CC data. For high redshift data, $\Omega_m$ is consistent with $\Omega_m = 0$, implying a constant Hubble parameter $H(z)=\textrm{constant} \sim 150$ km/s/Mpc. The secondary peak evident in the $H_0$ posterior arises from projecting the long tails in the $\Omega_m$ posterior onto the $H_0$ axis. The dashed lines correspond to the best fit parameters of the high redshift sample $(H_0, \Omega_m) = (150.37, 0.0098)$. This agrees well with the minimum $\chi^2$ from the MCMC chain $(H_0, \Omega_m) = (150.27,0.0099)$.}
\label{fig:CCsplit1} 
\end{figure}

Turning our attention from the best fits to the MCMC posteriors, we see that the central values corresponding to the peaks of $H_0$ and $\Omega_m$ posteriors track the best fits with reasonably good agreement through to $z_{\textrm{min}} = 0.6$. From $z_{\textrm{min}} = 0.7$ onwards we note that the minimum of the $\chi^2$ appears near the boundary of the $H_0$ posterior $68\%$ credible interval or beyond. Moreover, from the $\Omega_m$ posteriors, it is clear that there are long non-Gaussian tails in the direction of larger $\Omega_m$ values, which are in line with expectations when one fits the $\Lambda$CDM model to binned OHD at higher redshifts \cite{Colgain:2022tql}. Interestingly, at $z_{\textrm{min}} = 0.8$ and $z_{\textrm{min}} = 1$ (see Fig. \ref{fig:CCsplit1}), the $H_0$ posterior is bimodal with a secondary peak at smaller $H_0$ values. Increasing $z_{\textrm{min}}$ further, this secondary peak, which is an artefact of a projection effect, dominates (see later Fig. \ref{fig:CCzmin12}) and explains the discrepancy between the peak of the $H_0$ posterior and the best fit for $z_{\textrm{min}} = 1.2$ and $z_{\textrm{min}} = 1.4$. Evidently, the $H_0$ posterior is being pulled to lower values by a degeneracy in the 2D posterior. For $z_{\textrm{min}} = 1.5$ the $H_0$ peak and best fit agree well, but this is simply an accident due to a genuine low $H_0$ best fit.

{To reiterate, the disagreement between inferences from projected 1D MCMC posteriors and best fits is easily explained. Through to $z_{\textrm{min}} = 0.6$, the relatively good agreement observed is due to the 2D MCMC posterior being well constrained. From $z_{\textrm{min}} = 0.7$, the 2D posterior traces a banana-shaped contour in the $(H_0, \Omega_m)$-plane. The projection of this banana-shaped contour leads to the development of the bimodal distribution, where a genuine preference in the data for a larger $H_0$ best fit competes with a spurious $H_0$ peak due to a projection effect. From the Bayesian perspective, there is a degeneracy in the 2D posterior and this starts to impact MCMC inferences from $z_{\textrm{min}} = 0.7$ onwards, since this is where we start to see the equivalent of a $\sim 1 \sigma$ shift between the peak of the $H_0$ posterior and the best fit $H_0$ values. Nevertheless, in spite of the degeneracy, the $H_0$ posterior is attempting to track the $H_0$ best fit through to $z_{\textrm{min}} = 1$. This is evident through the increasing values of the $H_0$ posterior peak, but any signal is eventually washed out by the degeneracy between $H_0$ and $\Omega_m$. }

For the moment we leave physical speculations to the discussion and focus on the fact that CC data above $z=1$ favour less evolution in the Hubble parameter than the Planck-$\Lambda$CDM model. We would like to quantify the statistical significance of this trend, but since we are working in a {regime where a degeneracy is giving rise to non-Gaussian posteriors}, we can expect both Fisher matrix and MCMC to give biased results. In Fig.~\ref{fig:CCsplit1} we show MCMC posteriors for $z>1$ CC data in blue alongside posteriors for $z < 1$ CC data, which is simply added to aid comparison and also highlight the relative Gaussianity of the low redshift posteriors. One notes that the peaks of the $z > 1$ distributions are a little displaced from to the values minimising the $\chi^2$. However, the emergence of the lower peak in the $H_0$ posterior at $H_0 \sim 50$ km/s/Mpc has the hallmarks of a projection effect, and as promised, the $H_0$ posterior is bimodal. To appreciate the bimodality, note that the configurations in the blue banana-shaped curve in the top left corner of the 2D posterior are projected onto the lower $H_0$ peak. Moreover, if one shifts the $H_0$ peak from $H_0 \sim 150$ to $H_0 \sim 50$ km/s/Mpc while maintaining $\Omega_m \sim 0$, this shifts the magenta curve in Fig. \ref{fig:CC} outside of all the data points, so the lower $H_0$ peak is a phantom artefact unrelated to the goodness of fit. 

Ignoring these features, one could attempt to interpret the overlap in the 2D posteriors in Fig. \ref{fig:CCsplit1}. Doing so, one may conclude that low and high redshift CC data are consistent within $1 \sigma$. However, since Hubble tension is a 1D problem (local $H_0$ determinations are insensitive to other cosmological parameters), to compare with locally observed values of $H_0$ one needs to project onto the $H_0$ axis. Alternatively put, Hubble tension is a problem in 1D posteriors. {Projecting onto the $H_0$ axis, one finds that the degeneracy in the 2D posterior pulls the $H_0$ posterior to lower values. While this may be tolerable at $z_{\textrm{min}} = 1$, since the secondary peak is subdominant, as can be seen in Fig. \ref{fig:CCsplit1}, by $z_{\textrm{min}} = 1.2$ any preference the data has for a higher $H_0$ value has been washed out in the MCMC posteriors. This is clear  either from the $z_{\textrm{min}} = 1.2$ entry in Table \ref{tab:LCDM_CC} or from the later Fig. \ref{fig:CCzmin12}. The failure of MCMC to constrain the data can be contrasted with Fig. \ref{fig:CC} where it is obvious by the naked eye that one can fit an almost horizontal line through the CC data with $z > 1.2$. We return to this point later in section \ref{sec:MCMC_puzzle}.}

Evidently, given the non-Gaussian 1D posteriors, care is required when interpreting the significance of the trend towards a non-evolving (horizontal) $H(z)$ at higher redshifts in Fig.~\ref{fig:CC}. We cannot use the errors from the Fisher matrix as we are clearly in a non-Gaussian regime, whereas MCMC inferences are impacted by projection effects. For this reason, we resort to mock simulations. While this may seem a little redundant if we are going to employ profile distributions in section \ref{sec:PD}, there is motivation for this exercise. In \cite{Colgain:2022rxy} the significance of a descending $H_0$/increasing $\Omega_m$ trend with effective redshift in OHD, Type Ia SNe and QSOs was estimated to be a $\sim 3 \sigma$ effect on the basis of combining $\sim 2 \sigma$ effects in each of the \textit{independent} data sets using Fisher's method. Here, working with the same data throughout, we can directly compare the significance of a discrepancy estimated through mock simulations from the significance of a discrepancy estimated through profile distributions. In particular, we will address the question: how significant is a constant $H(z)$ with $z_{\textrm{min}}=1$ (8 data points) against the Planck consistent cosmology favoured by the full data set ($z_{\textrm{min}}=0$ entry in Table \ref{tab:LCDM_CC})? Note, the significance will be overestimated due to missing systematic uncertainties (see \cite{Moresco:2020fbm}), but we can still make comparison between mock simulations and profile distributions. 

\subsection{Mock Simulations}
\begin{figure}[htb]
   \centering
\includegraphics[width=76.7mm]{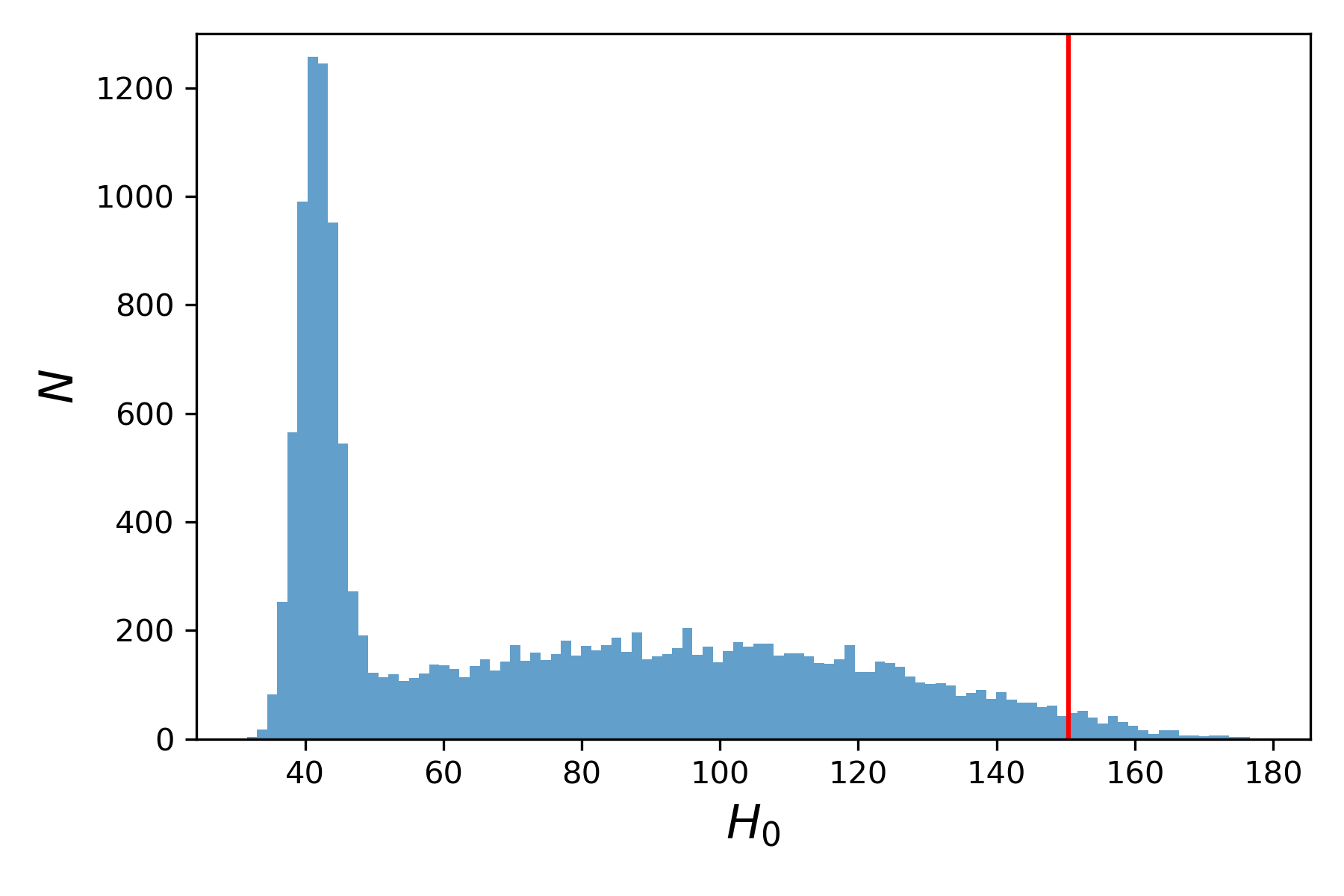}  
\includegraphics[width=76.7mm]{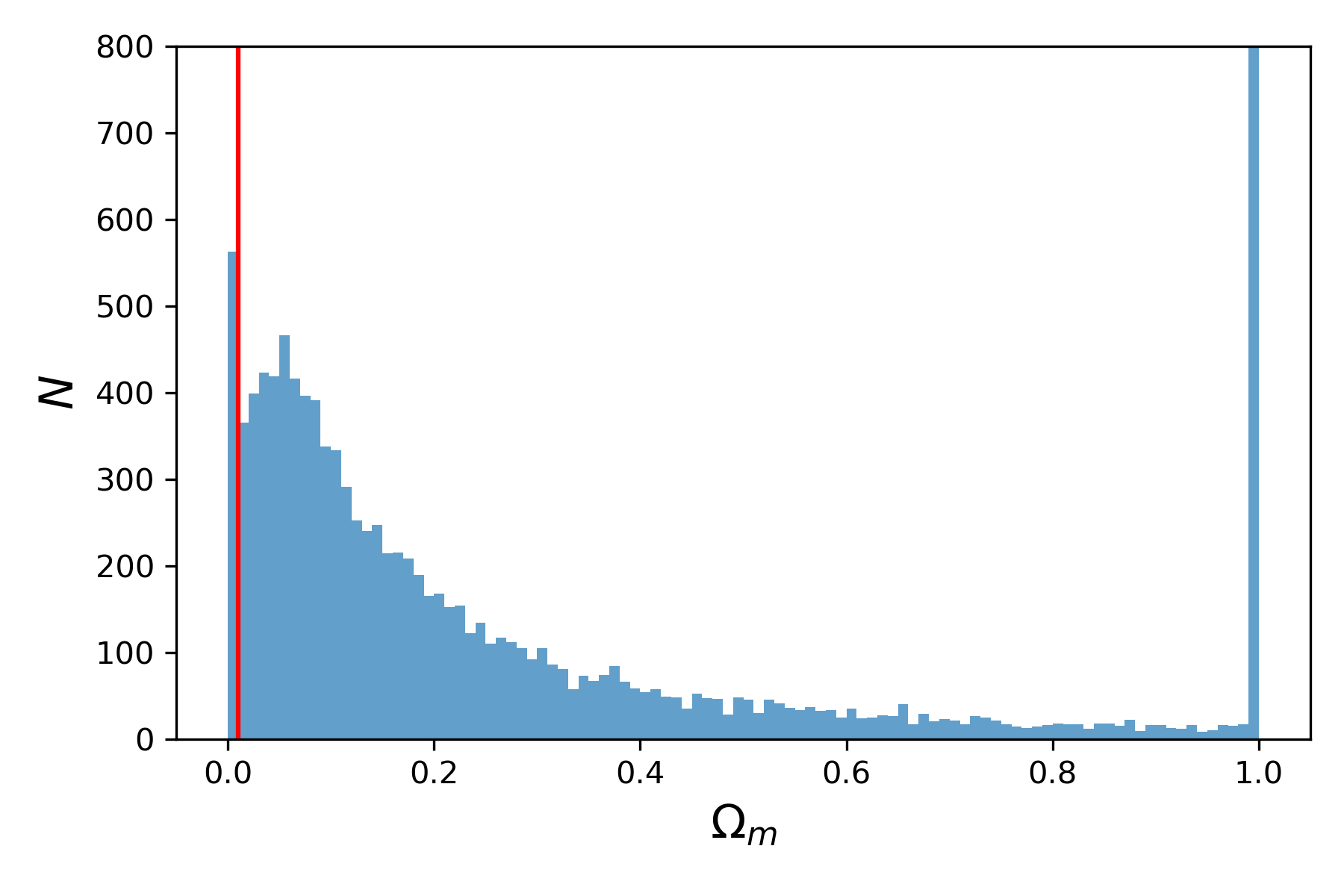}
\caption{PDFs of best fit parameters from mock CC data with $z_{\textrm{min}}=1$ (8 data points). The injected $(H_0, \Omega_m)$ parameters come from the MCMC chain for the full sample ($z_{\textrm{min}} = 0$ entry in Table \ref{tab:LCDM_CC}). The best fits from observed data are denoted by solid red lines. Numerous best fits saturate our bounds both at $\Omega_m=0$ and $\Omega_m =1$ leading to an unusual peak in the $H_0$ PDF at $H_0 \sim 40$ km/s/Mpc. Removing this feature, it is evident that the $H_0$ peak and $\Omega_m$ peak have been shifted to larger and smaller values, respectively, in line with expectations from a projection effect. There is a low probability ($p = 0.022$ for $H_0$, $p=0.035$ for $\Omega_m$) of finding larger $H_0$ and smaller $\Omega_m$ than observed best fits (red lines) from the mock best fits.}
\label{fig:CCsims} 
\end{figure}

To address the question of statistical significance using mock simulations, we begin with the MCMC chains for the full sample. For each entry in the MCMC chain (approximately 15,000 entries in total), we generate a new realisation of the 8 high redshift data points $(z > 1)$ that are by construction statistically consistent with both the best fits from the full sample and also the Planck-$\Lambda$CDM values \cite{Planck:2018vyg}. More concretely, for each $(H_0, \Omega_m)$ entry in our MCMC chain, we displace the data points to the corresponding $\Lambda$CDM Hubble parameter before generating new data points in a normal distribution where the errors serve as standard deviations. We then fit back the $\Lambda$CDM model to each realisation of the mock data and record the best fit $(H_0, \Omega_m)$ values, which give us a distribution of expected $(H_0, \Omega_m)$ best fits. The distributions are presented in Fig.~\ref{fig:CCsims} alongside the best fits from observed data. Throughout, we assume canonical values $(H_0, \Omega_m) = (70, 0.3)$ for the initial guess of the fitting algorithm. Best fits can saturate our bounds, i. e. $\Omega_m = 0$ and $\Omega_m = 1$, and this leads to an unsightly pile up of best fits at $\Omega_m = 0$ and $\Omega_m = 1$ in Fig.~\ref{fig:CCsims}. It is important to retain all the configurations, otherwise one is not accounting for the probability that a best fit falls outside our priors. As a consistency check, we see that the median or 50$^{\textrm{th}}$ percentile, $(H_0, \Omega_m) = (68.32, 0.321)$ agrees well with the mock input parameters, thereby demonstrating that there are an equal number of best fits with values above and below the injected parameters in the mocks. We find that probability of a more extreme (larger) $H_0$ value to be $p = 0.022$, while the probability of a more extreme (smaller) $\Omega_m$ value to be $p = 0.035$, respectively. Converted into a Gaussian statistic, these correspond to $2 \sigma$ and $1.8 \sigma$, respectively, for a one-sided normal distribution. Thus, on the basis of mock simulations, we estimate the non-evolving constant $H(z)$ with $z_{\textrm{min}} = 1$ as a $\sim 2 \sigma$ effect. In the next section we will recover this number more or less from the profile distribution analysis. 

\section{Profile Distributions}
\label{sec:PD}
Having explained the mathematics behind the bias, which gives rise to a projection effect, in subsection \ref{sec:math}, and having illustrated how it affects MCMC inferences in subsection \ref{sec:CCbias} - the minimum of the $\chi^2$ risks falling outside of $68\%$ credible intervals - we turn to profile distributions (PDs) \cite{Gomez-Valent:2022hkb}, an extension of the profile likelihood, e. g. \cite{Trotta:2017wnx}, to provide a complementary viewpoint. Consider two sets of parameters $\theta_1$ and $\theta_2$ and a normalised distribution $\mathcal{P}(\theta_1, \theta_2)$. The basic idea \cite{Gomez-Valent:2022hkb} is to study the ratio 
\be
\label{R}
R(\theta_1) = \frac{\tilde{\mathcal{P}}(\theta_1)}{\max_{\theta_1} \tilde{\mathcal{P}}(\theta_1) } = \frac{\tilde{\mathcal{P}}(\theta_1)}{\max_{\theta_1, \theta_2} \mathcal{P}(\theta_1, \theta_2) },  
\ee
where $\tilde{\mathcal{P}}(\theta_1)$ is the PD, defined to be the maximum of $\mathcal{P}$ for each $\theta_1$ along the $\theta_2$ direction: 
\be
\label{PD}
\tilde{\mathcal{P}} (\theta_1) = \max_{\theta_2} \mathcal{P}(\theta_1, \theta_2). 
\ee
The advantage of this approach is that $R(\theta_1)$ can serve as a probability distribution function (up to an overall normalisation), however we do not need to perform any integration, so $R(\theta_1)$ is not prone to volume or projection effects. {It is worth noting that $R(\theta_1)$ is a profile likelihood ratio in more common terminology \cite{Trotta:2017wnx}.} At this juncture, given the simplicity of our setup with only two parameters $(H_0, \Omega_m)$, we can be more explicit. Consider the probability distribution,   
\be
\mathcal{P}(\theta_1, \theta_2) = \exp \left( - \frac{1}{2} \chi^2(\theta_1, \theta_2) \right), 
\ee
where $\theta_i \in \{H_0, \Omega_m \}$  and $\chi^2(H_0, \Omega_m)$ is our earlier likelihood (\ref{eq:chi2}). The maximum value of $\mathcal{P}$ occurs for the minimum value of $\chi^2$ from the MCMC chain, $\mathcal{P}_{\textrm{max}} = e^{-\frac{1}{2} \chi^2_{\textrm{min}}}$. In this concrete setting, the PD becomes 
\be
\tilde{\mathcal{P}}(\theta_1) = e^{-\frac{1}{2} \chi^2_{\textrm{min}}(\theta_1)}, 
\ee
where $\chi^2_{\textrm{min}}(\theta_1)$ denotes the minimum value of the $\chi^2$ along the $\theta_2$ direction for a fixed $\theta_1$ value. It should not be confused with the overall minimum $\chi^2_{\textrm{min}}$, which can be extracted easily from the MCMC chain. In practice, one can also determine $\chi^2_{\textrm{min}}(\theta_1)$ from the MCMC chain by breaking the $\theta_1$ direction up into bins and finding the minimum of the $\chi^2$ for each bin. Having done so, we are in a position to define a PDF \cite{Gomez-Valent:2022hkb}: 
\be
\label{eq:w}
w(\theta_1) = \frac{e^{-\frac{1}{2} \chi^2_{\textrm{min}}(\theta_1)}}{\int e^{-\frac{1}{2} \chi^2_{\textrm{min}}(\theta_1)} \, \textrm{d} \theta_1} = \frac{R(\theta_1)}{\int R(\theta_1) \, \textrm{d} \theta_1}, 
\ee
where in the second equality we have divided top and bottom by $\mathcal{P}_{\textrm{max}} = e^{-\frac{1}{2} \chi^2_{\textrm{min}}}$. As a result, $R(\theta_1) = e^{-\frac{1}{2} \Delta \chi_{\textrm{min}}^2}$, where $\Delta \chi^2_{\textrm{min}} := \chi_{\textrm{min}}^2(\theta_1) - \chi^2_{\textrm{min}}$, so that $R(\theta_1)$ peaks at $R(\theta_1) = 1$. Note that $\int_{-\infty}^{+\infty} w(\theta_1) \, \textrm{d} \theta_1 = 1$ by construction, so $w(\theta_1)$ describes a properly normalised PDF. Thus we can identify the $1 \sigma, 2 \sigma$ and $3 \sigma$ confidence intervals corresponding to the 68\%, 95\% and 99.7\% confidence level, respectively, by simply identifying $\theta_1^{(1)}$ and $\theta_1^{(2)}$ such that \cite{Gomez-Valent:2022hkb}
\be
\label{eq:wsigma}
\int_{\theta_1^{(1)}}^{\theta_1^{(2)}} w(\theta_1) \, \textrm{d} \theta_1 = I, \ \ w(\theta_1^{(1)}) = w(\theta_1^{(2)}), \ \ I \in \{0.68, 0.95, 0.997\}. 
\ee
We will outline how these conditions can most easily be satisfied when we turn to explicit examples. 

Our first port of call is making sure that the PD methodology gives sensible results. This can be best judged by applying it to CC data with $z_{\textrm{min}} = 0$, since this is where we expect a distribution closest to a Gaussian distribution, as is evident from the agreement between Fisher matrix and MCMC results in Table \ref{tab:LCDM_CC}. In particular, we will be interested in a comparison between $68\%$ ($1 \sigma$) confidence intervals to make sure that (\ref{eq:wsigma}) is not underestimating or overestimating the confidence interval. 

\begin{figure}[htb]
   \centering
\includegraphics[width=76.7mm] {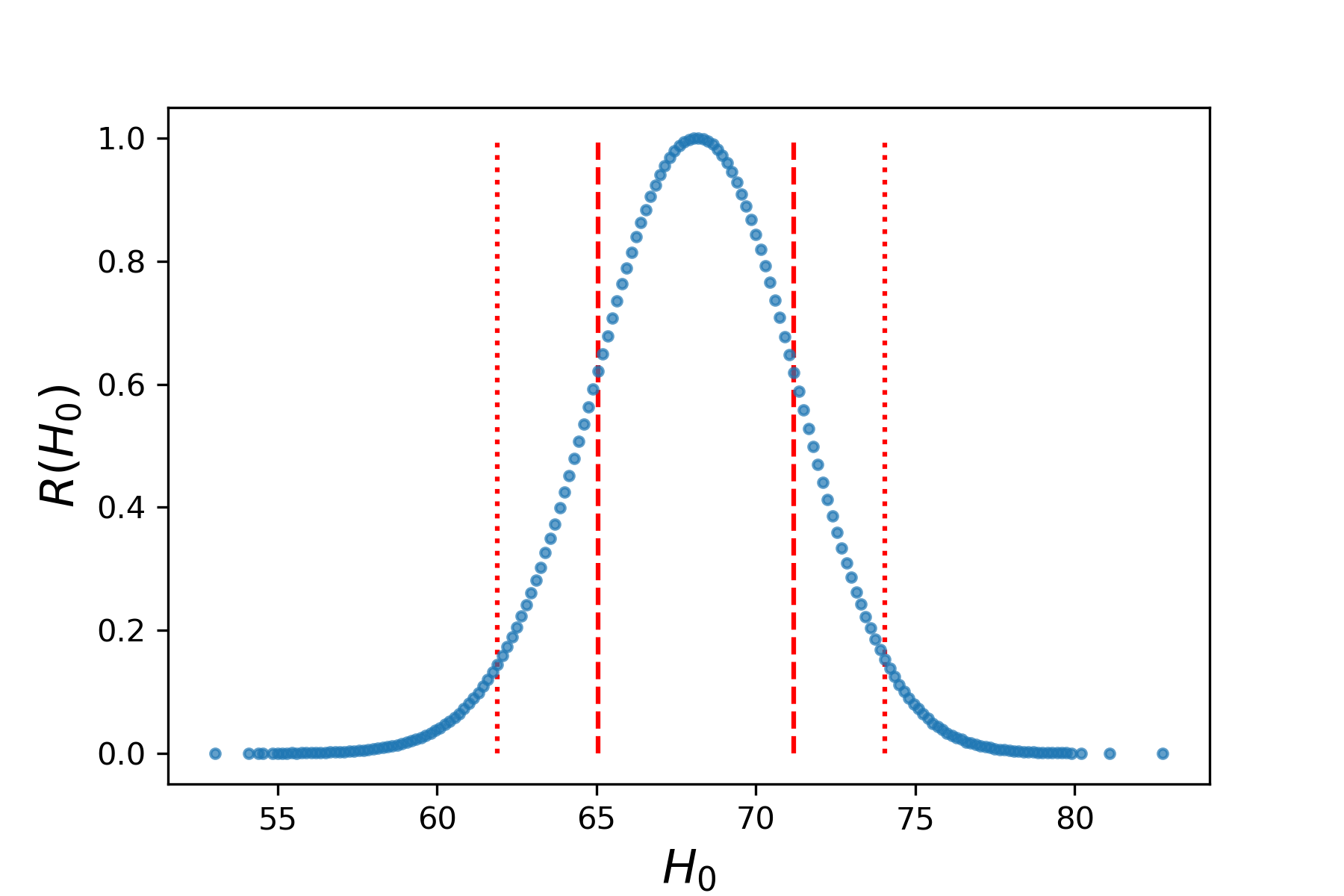} 
\includegraphics[width=76.7mm] {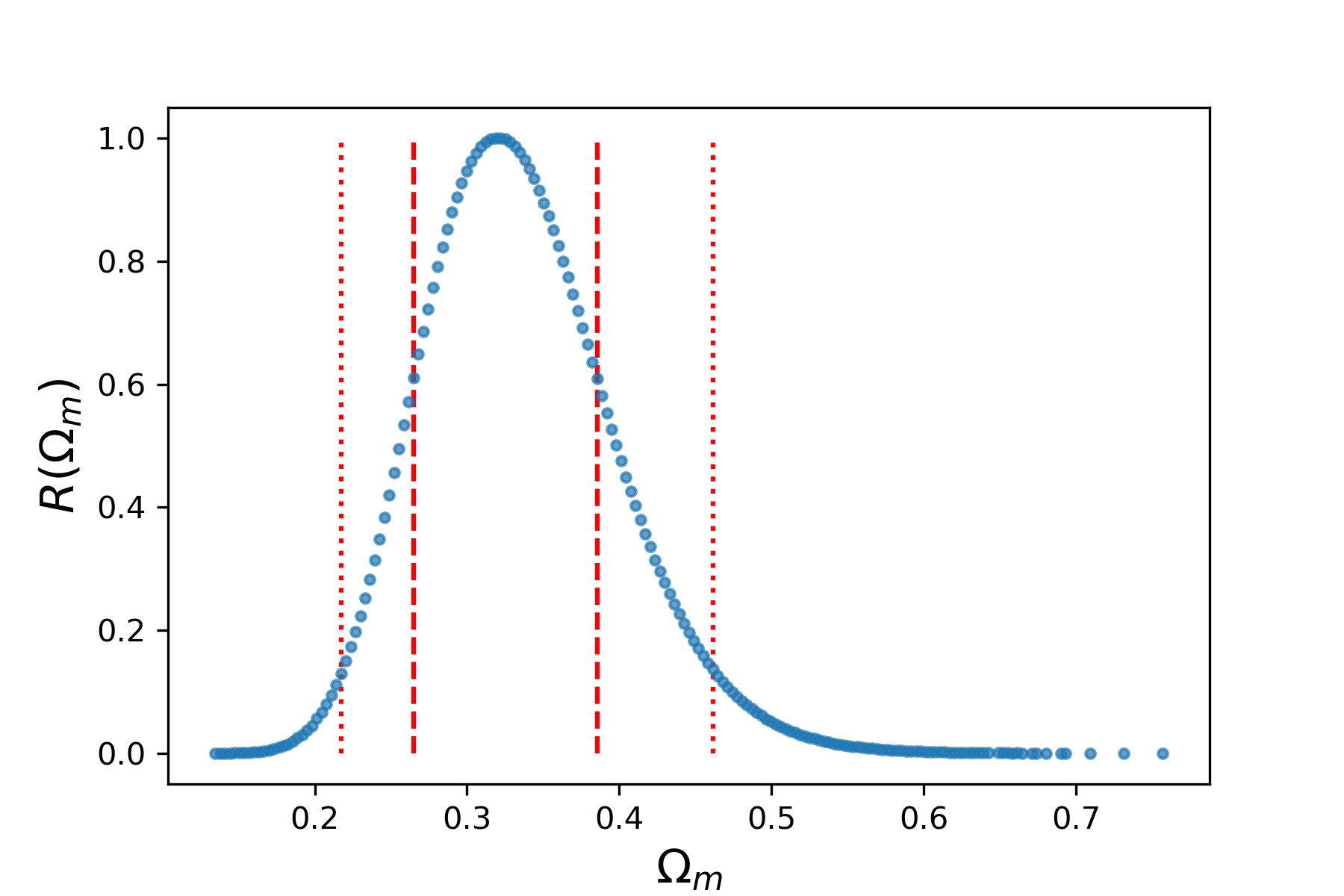}
\caption{$R(H_0)$ and $R(\Omega_m)$ distributions for the full CC data set ($z_{\textrm{min}} = 0$) in Fig.~\ref{fig:CC}. The dashed and dotted lines denote $68\%$ and $95\%$ confidence intervals.}
\label{fig:R_zmin0}
\end{figure}

We start by running a long MCMC chain (100,000 iterations) in order to ensure bins are well populated, and begin by analysing $\theta_1 = H_0$ with $\theta_2 = \Omega_m$. From the MCMC chain we identify the smallest and largest value of $H_0$ in the chain and break up this range into approximately 200 uniform bins, which we label using the $H_0$ value at the centre of the bin. We omit any empty bins. One can increase the number of bins by simply running a longer MCMC chain. In each $H_0$ bin we identify the minimum value of the $\chi^2$, $\chi^2_{\textrm{min}}(H_0)$, and calculate $R(H_0)$. One then repeats the steps for $\Omega_m$. In Fig.~\ref{fig:R_zmin0} we plot $R(H_0)$ against $H_0$ and $R(\Omega_m)$ against $\Omega_m$, noting that the distributions are Gaussian to first approximation. 

Since the distributions from the MCMC chain are sparse in the tails, empty bins are evident in Fig.~\ref{fig:R_zmin0}. Nevertheless, with 200 bins, modulo any empty bins, we have sufficient density of points to calculate the total area under the $R(H_0)$ and $R(\Omega_m)$ curve using Simpson's rule. Any concern about precision can simply be mitigated by running a longer MCMC chain and increasing the number of bins. 
One may directly use $R(H_0)\leq 1$ and $R(\Omega_m)\leq 1$   to find $68$, $95$ and $99.7$ percentiles,  respectively corresponding to $1 \sigma, 2 \sigma$ and $3 \sigma$ confidence intervals. Consider $F_\kappa:= \int_{R\geq \kappa} R (\theta_1) \, \textrm{d} \theta_1$, where $\kappa \leq 1$. Observe that $F_{\kappa=1}=0$ and $F_{\kappa=0}:=F_0=\int R(\theta_1) \textrm{d} \, \theta_1$. Then move $\kappa$ through and terminate the process when $F_\kappa/F_0$ is equal to $0.68$, $0.95$ and $0.997$. This gives the corresponding range for $\theta_1$ that defines the confidence interval.
Working with the precision afforded to us by approximately 200 bins, the $H_0$ and $\Omega_m$ $68\%$ confidence intervals are presented in Fig.~\ref{fig:R_zmin0} and the first entry in Table \ref{tab:LCDM_CC_PD}. The outcome is in excellent agreement with both Fisher matrix and MCMC analysis. In particular, a mild non-Gaussianity in $\Omega_m$ is evident in both Fig.~\ref{fig:R_zmin0} and the errors. 
Thus, we have succeeded in recovering results in the (almost) Gaussian regime that are consistent with Fisher matrix and MCMC analysis and this provides an important check of the methodology. 

{As further remarks, we note that when profile likelihoods are Gaussian (see discussion in \cite{Colgain:2024clf}), as is visibly the case here in Fig.~\ref{fig:R_zmin0}, it is valid to apply Wilks' theorem \cite{Wilks:1938dza}. Doing so, $68 \%$ confidence intervals correspond to regions with $\Delta \chi^2 \leq 1 \Leftrightarrow R(\theta_1) \gtrsim 0.6$. One can now draw a horizontal line across from $R(\theta_1) \sim 0.6$, $\theta_1 \in \{H_0, \Omega_m \}$, in Fig.~\ref{fig:R_zmin0} and confirm that the line intersects $R(\theta_1)$ at the dashed lines demarcating $68\%$ confidence intervals. The key point here is that despite the methodology in \cite{Gomez-Valent:2022hkb} being new, the resulting confidence intervals agree with standard methods when distributions are Gaussian. It should be noted that \cite{Gomez-Valent:2022hkb} bins the MCMC chain to construct PDs, which are essentially profile likelihood ratios, but this is acceptable, as explained in \cite{Trotta:2017wnx}. As shown in Fig. 4 of \cite{Colgain:2024clf}, for converged MCMC chains, the agreement between profile likelihoods constructed through optimisation and profile likelihoods constructed through binned MCMC chains is good. Here, in contrast to \cite{Colgain:2024clf}, we run very long chains, well beyond the point at which they converge, and our model has fewer nuisance parameters, so the agreement is expected to be excellent. 

\begin{figure}[htb]
   \centering
\includegraphics[width=76.7mm] {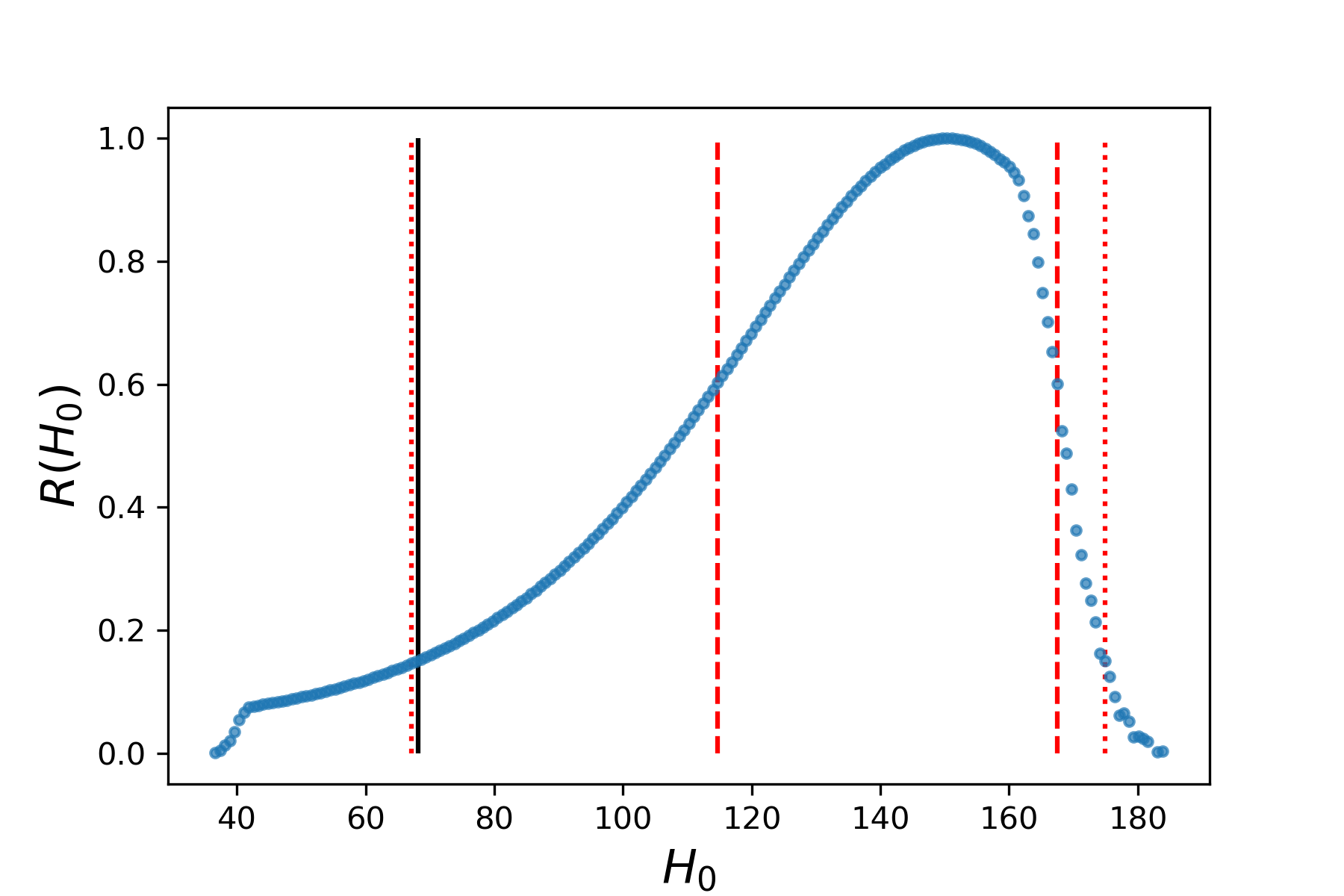} 
\includegraphics[width=76.7mm] {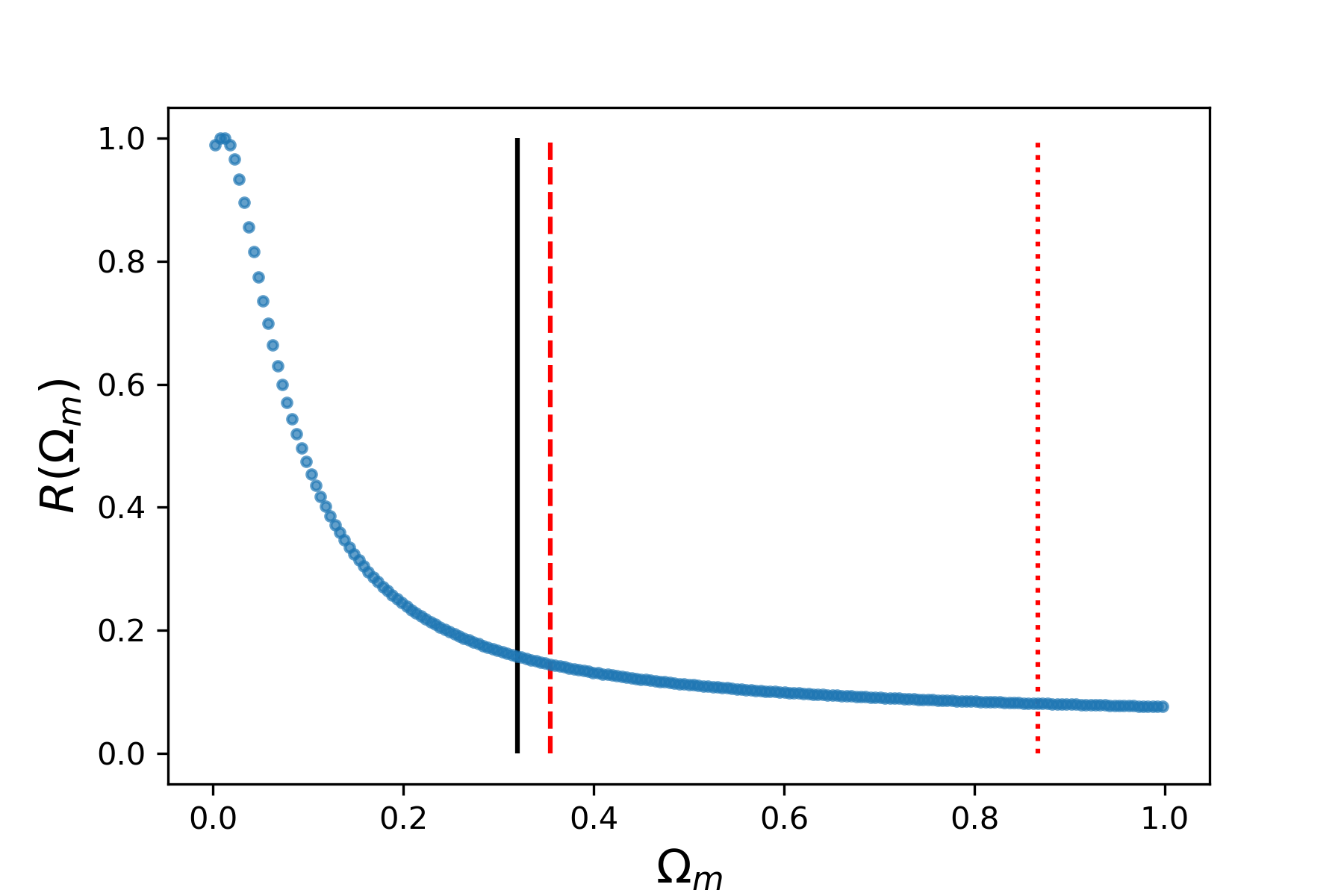}
\caption{Same as Fig.~\ref{fig:R_zmin0} but a truncated CC data set of 8 high redshift data points with $z_{\textrm{min}}=1$. The black lines denote the best fit values from the full sample $(H_0, \Omega_m) = (68.2, 0.320)$.}
\label{fig:R_zmin1} 
\end{figure}

We now apply the PD methodology to the non-Gaussian regime where MCMC marginalisation leads to biased results {and Wilks' theorem \cite{Wilks:1938dza} is no longer valid}. To be concrete, we focus on the eight data points in the range $1 < z < 2$ where a non-evolving $H(z)$ trend is evident in the raw data in Fig.~\ref{fig:CC}. Our goal here is to quantify the disagreement with the full data set, where one infers $H_0 \sim 68.2$ km/s/Mpc and $\Omega_m \sim 0.32$. A similar exercise was performed in subsection \ref{sec:features} with mock simulations and the disagreement was estimated to be approximately $2 \sigma$. Repeating the steps outlined above for the CC data with $z_{\textrm{min}} = 1$ we find the distributions in Fig.~\ref{fig:R_zmin1}. The first observation is that the distributions are non-Gaussian, but a comparison to the MCMC posteriors from the same data in blue in Fig.~\ref{fig:CCsplit1} reveals that there is no secondary $H_0$ peak at $H_0 \sim 50$ km/s/Mpc. Thus, we confirm the secondary peak to be a projection effect. That being said, the primary $H_0$ peak from Fig.~\ref{fig:CCsplit1} has shifted to the dashed line corresponding to the minimum of the $\chi^2$, since the peak of the $R(H_0)$ distribution and $\chi^2$ minimum agree by construction. Comparing the blue $\Omega_m$ distribution from Fig.~\ref{fig:CCsplit1} to the $R(\Omega_m)$ distribution in Fig.~\ref{fig:R_zmin1}, we see that the peak is close to $\Omega_m = 0$ and that the tails continue to $\Omega_m = 1$. In both plots we see that there is a non-zero probability of inferring $\Omega_m = 1$. In some sense, this is not so surprising, the reason being that one is free to adopt generous priors for $H_0$, so that probability of large and small $H_0$ values is zero, but the priors on $\Omega_m$ in the flat $\Lambda$CDM model are restricted. For this reason, as a distribution spreads one invariably finds that distributions are impacted by the $\Omega_m$ priors.\footnote{Note, this is a problem for the flat $\Lambda$CDM model. In particular, one may easily find that the peak of the $\Omega_m$ distribution is larger than $\Omega_m=1$, as is the case with Hubble Space Telescope SNe with redshifts $z > 1$ in the Pantheon+ sample \cite{Malekjani:2023dky}.}

It is evident from Fig.~\ref{fig:R_zmin1} that any tension that exists is confined to the $H_0$ parameter. Moreover, since there may be only one binned value of $\Omega_m$ below the $R(\Omega_m)$ peak, at the precision afforded to us by 200 bins, the $R(\Omega_m)$ distribution in Fig.~\ref{fig:R_zmin1} is essentially one-sided and the $68\%$ confidence interval stretches beyond $\Omega_m \sim 0.32$, so there is no disagreement in the $\Omega_m$ parameter. Nevertheless, in the $H_0$ parameter we see that $H_0 \sim 68.2$ km/s/Mpc, the value favoured by the full data set is at the boundary of the $95\%$ confidence interval, i.e. just under $2 \sigma$ removed from the peak. The main point here is that, as is obvious from the raw data, current CC data with $z > 1$ has a preference for a non-evolving Hubble parameter $H(z)$ with a large constant $H_0 \sim 150$ km/s/Mpc. The disagreement is just under $2 \sigma$, more accurately a $94\%$ ($1.9 \sigma$) confidence level difference from $R(H_0)$, and only $63\%$ ($0.9 \sigma$) confidence level difference from $R(\Omega_m)$. 

Note, since $R(\Omega_m)$ is curtailed by the bounds on $\Omega_m$, inferences based on $\Omega_m$ need to be taken with a pinch of salt as they depend on our bounds. Moreover, from $R(H_0)$ we note that the $1^{\textrm{st}}$ derivative fails to be continuous at $H_0 \sim 40$ km/s/Mpc, which also appears to be an artefact of the $\Omega_m \leq 1$ bound. This kink happens beyond the $95 \%$ confidence intervals, so even if corrected by relaxing the bounds on $\Omega_m$ it is not expected to greatly impact the results. We note that while it is not correct to apply Wilks' theorem, the $R(H_0) \sim 0.6$ horizontal line appears to intersect $R(H_0)$ at the dashed lines denoting $68 \%$ confidence intervals. This is not the case with $R(\Omega_m)$, where the bounds are impacting results. We note that one can use the Feldman-Cousins prescription \cite{Feldman:1997qc} to generalise methodology based on Wilks' theorem when boundaries are present, but once again the methodology is only valid for Gaussian profiles, which is demonstrably not the case here. Ultimately, the methods in the literature rest upon Gaussian distributions being recovered in the large sample limit. In cosmology, whether one is in the large sample limit or not can only be judged once one has constructed the MCMC posterior or profile likelihood. Blindly applying theorems beyond their scope is neither good mathematics nor good physics.

\begin{figure}[htb]
   \centering
\includegraphics[width=90mm]{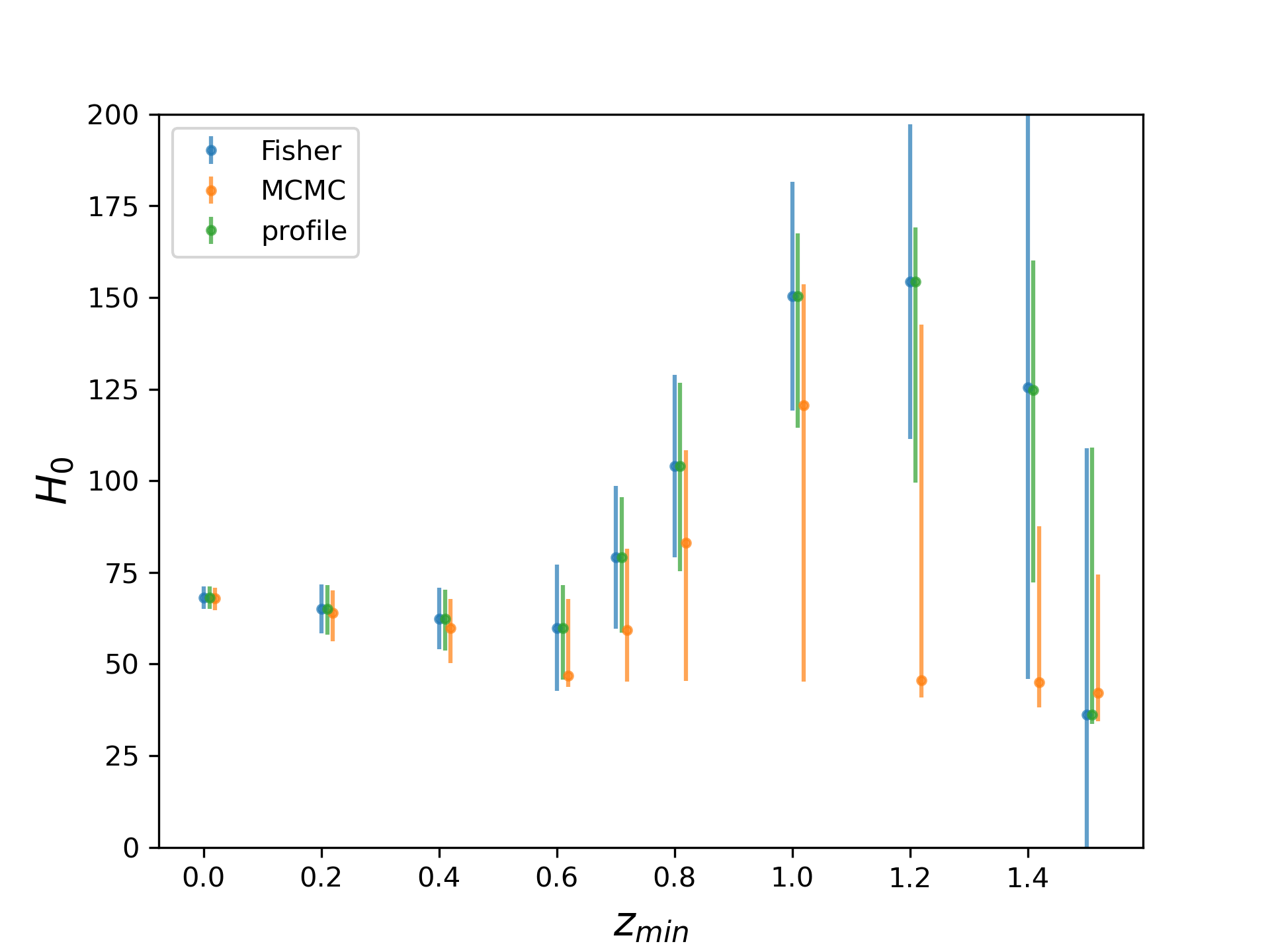}
\caption{$H_0$ constraints versus $z_{\textrm{min}}$ from Table \ref{tab:LCDM_CC} and Table \ref{tab:LCDM_CC_PD}. MCMC constraints follow Fisher matrix and profile constraints well at $z_{\textrm{min}} \lesssim 0.6$ and at $z_{\textrm{min}} = 1.5$, but the credible intervals shift with respect to the confidence intervals to lower values at intermediary $z_{\textrm{min}}$. For presentation purposes we have shifted the constraints in $z_{\textrm{min}}$.}
\label{fig:H0trend} 
\end{figure}

Although $\sim 2 \sigma$ may not be a serious discrepancy, essentially because of the poor data quality (8 data points), this disagreement supports the $\sim 2 \sigma$ discrepancy seen in the mock simulations. It should be borne in mind that systematic uncertainties have been omitted and these will reduce this discrepancy once properly propagated. Given the agreement between the PD and mock simulation analysis, there is nothing to suggest that the three independent trends highlighted in \cite{Colgain:2022rxy} across OHD, Type Ia SNe and QSOs are not \textit{bona fide} disagreements and that redshift evolution is present in the sample. The task remains to combine them at the level of a $\chi^2$ likelihood instead of combining them using Fisher's method on the basis that they are independent probabilities. We leave this exercise for a forthcoming paper, but revisit the tension in OHD data in section \ref{sec:tension}. 
For completeness, in Table \ref{tab:LCDM_CC_PD} we perform a reanalysis of CC data subsets with the PD approach and record the $68 \%$ confidence intervals. In Fig. \ref{fig:H0trend} we compare the $H_0$ constraints from Table \ref{tab:LCDM_CC} and Table \ref{tab:LCDM_CC_PD}, where it is clear that MCMC credible intervals shift to lower values at intermediate $z_{\textrm{min}}$. 

\begin{table}
    \centering
    \begin{tabular}{|c|c|c|c|}
    \hline
    \rule{0pt}{3ex} $z_{\textrm{min}}$ & \# CC & \multicolumn{2}{|c|}{PD}  \\
    \hline
    \rule{0pt}{3ex} & & $H_0$ (km/s/Mpc) & $\Omega_m$ \\
    \hline\hline
    \rule{0pt}{3ex} $0$ & $34$ & $68.15^{+3.04}_{-3.11}$ & $0.320^{+0.065}_{-0.055}$ \\
    \hline 
    \rule{0pt}{3ex} $0.2$ & $27$ & $65.03^{+6.52}_{-7.03}$ & $0.368^{+0.167}_{-0.110}$ \\
    \hline 
    \rule{0pt}{3ex} $0.4$ & $22$ & $62.42^{+7.78}_{-8.74}$ & $0.411^{+0.236}_{-0.113}$ \\
    \hline
    \rule{0pt}{3ex} $0.6$ & $15$ & $59.75^{+11.73}_{-13.97}$ & $0.455^{+0.355}_{-0.160}$ \\
    \hline
    \rule{0pt}{3ex} $0.7$ & $14$ & $79.10^{+16.42}_{-20.56}$ & $0.222^{+0.386}_{-0.117}$ \\
    \hline
    \rule{0pt}{3ex} $0.8$ & $11$ & $103.94^{+22.88}_{-28.54}$ & $0.097^{+0.378}_{-0.074}$ \\
    \hline
    \rule{0pt}{3ex} $1$ & $8$ & $150.35^{+17.12}_{-35.95}$ & $ < 0.339$ \\
    \hline
    \rule{0pt}{3ex} $1.2$ & $7$ & $154.26^{+14.88}_{-54.82}$ & $ < 0.570$ \\
    \hline
    \rule{0pt}{3ex} $1.4$ & $4$ & $124.81^{+35.38}_{-52.60}$ & $ < 0.661$ \\
    \hline
    \rule{0pt}{3ex} $1.5$ & $3$ & $36.11^{+72.87}_{-2.43}$ & $ > 0.354$\\
    \hline
    \end{tabular}
    \caption{Same as Table \ref{tab:LCDM_CC} but with the PD methodology in lieu of Fisher matrix and MCMC analysis. The high redshift $R(\Omega_m)$ distributions are typically one-sided, so one encounters $68\%$ confidence interval upper and lower bounds.}
    \label{tab:LCDM_CC_PD}
\end{table}

\section{{An MCMC puzzle}}
\label{sec:MCMC_puzzle}
It should be evident that a degeneracy (banana-shaped contour) in the 2D MCMC posterior, which gives rise to a projection effect in 1D posteriors, explains differences between posterior peaks and best fits in Table \ref{tab:LCDM_CC}. Moreover, the effect of this degeneracy starts to impact results at relatively low redshifts, i. e. $z_{\textrm{min}} = 0.7$. More puzzling still, as is clear from Fig. \ref{fig:CCzmin12}, by $z_{\textrm{min}}= 1.2$ any preference in the data for a higher $H_0$ value has been lost \footnote{{We use \textit{emcee} on its default algorithm. It is plausible that changing the algorithm has a bearing on the outcome, but it is not expected to remove the degeneracy, since the $\Lambda$CDM model must transition to an effective 1D model in high redshift bins. This underscores Bayesian cosmology intuition that high redshift data only constrains the combination $\Omega_m h^2$. The point here is that intuition rests upon experience with MCMC, so Fig. \ref{fig:CCzmin12} is expected regardless of the algorithm.}}. All that remains is a bump in the $H_0$ posterior that coincides with the best fit, $H_0 = 154.35$ km/s/Mpc (dashed line). Nevertheless, returning to the raw data, it is blatantly clear that one can fit an almost horizontal line to CC data with $z > 1.2$. Moreover, one can get such a function from the $\Lambda$CDM model (\ref{eq:lcdm}) through a small value of $\Omega_m$ and a large value of $H_0$, and there is no reason to believe that the data poorly constrains the model. Nevertheless, this is the conclusion a cosmologist reaches if they rely exclusively on MCMC marginalisation for their ground truth. Note, we do not need to construct a profile likelihood or distribution to arrive at this conclusion; it suffices to visibly inspect the data and understand that the $\Lambda$CDM model reduces to $H(z) = \textrm{constant}$ in the limit $\Omega_m \rightarrow 0$.

\begin{figure}[htb]
   \centering
\includegraphics[width=80mm]{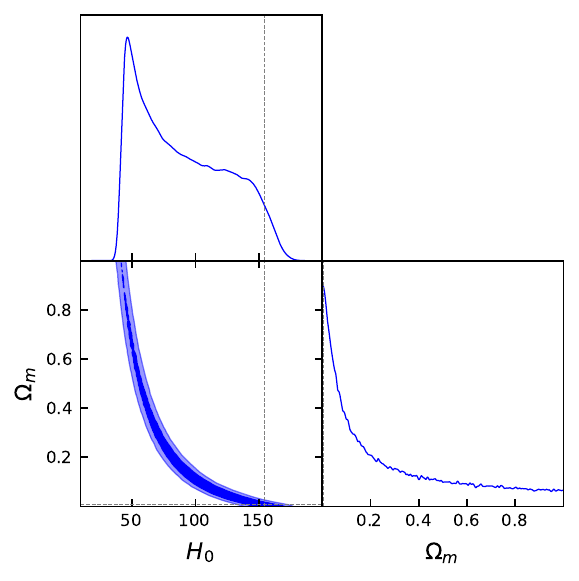}
\caption{{Cosmological parameter inferences from MCMC chains for the CC data with $z > 1.2$. Dashed lines denote the minimum of the $\chi^2$. 1D $H_0$ posteriors are impacted by a degeneracy in the 2D posterior.}}
\label{fig:CCzmin12} 
\end{figure}

{Taking this further, and leaving aside the obvious trend in the data for the moment, if the degeneracy in the 2D MCMC posterior is physical, one expects it to translate into an almost constant $\chi^2$ curve in the $(H_0, \Omega_m)$-plane. In essence, there is still a $\chi^2$ minimum, but it should be part of a curve where changes in $\chi^2$ relative to the minimum are minimal. This can be tested through PD analysis. In Fig. \ref{fig:R_zmin12} we show $R(H_0)$ and $R(\Omega_m)$ distributions, where it is evident that noticeable changes in $\chi^2$ occur as we move away from the minimum of the $\chi^2$. We find that the black lines corresponding to the best fits of the full sample ($z_{\textrm{min}} = 0$) occur at $87 \%$ ($1.5 \sigma$) confidence level in $R(H_0)$ and $47 \%$ ($0.6 \sigma$) confidence level in $R(\Omega_m)$. Our PD analysis essentially supports what one sees by naked eye in the raw data; there \textit{is} enough constraining power in the data to confirm the preference for an almost constant Hubble parameter.}

\begin{figure}[htb]
   \centering
\includegraphics[width=76.7mm] {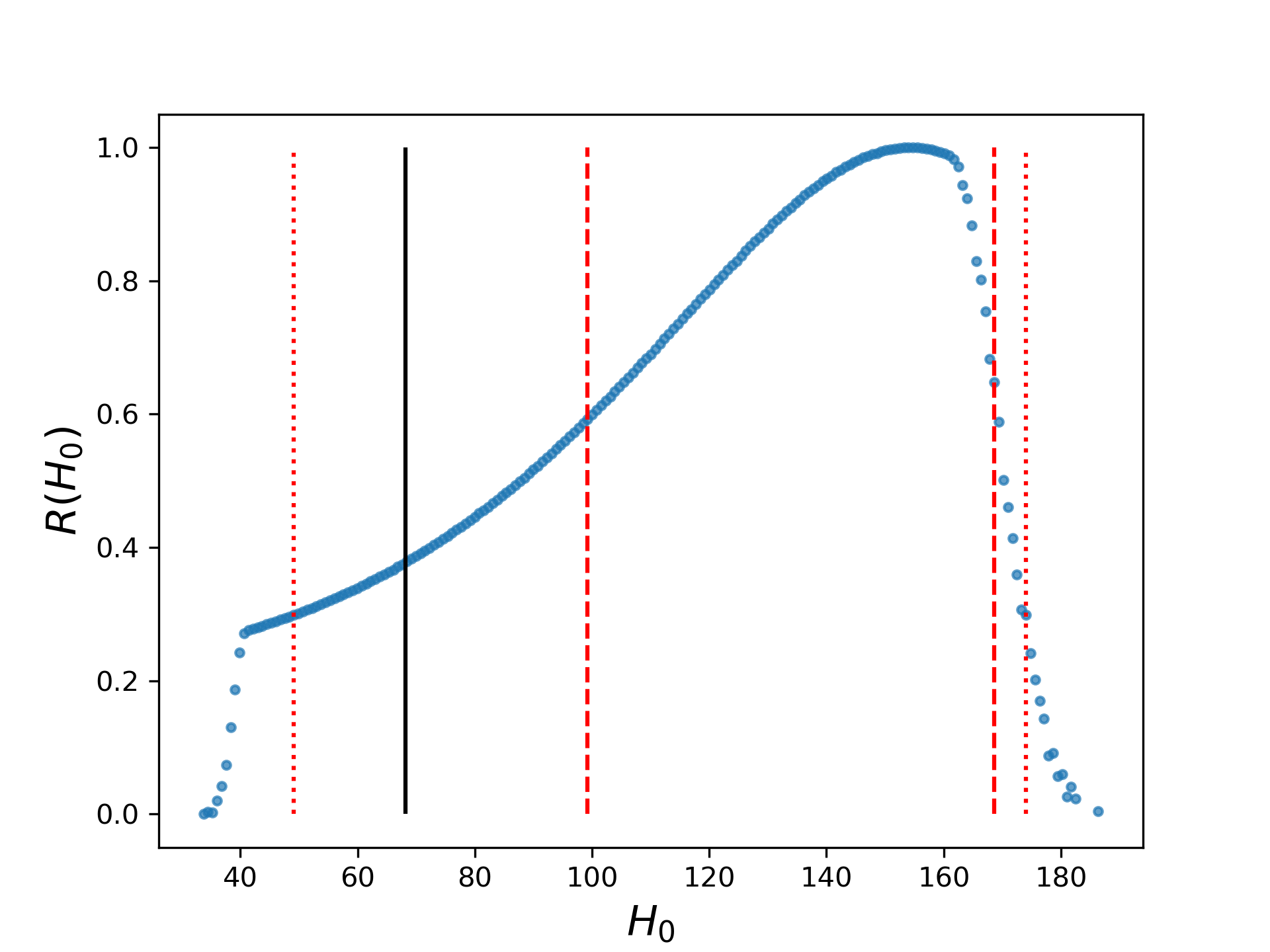} 
\includegraphics[width=76.7mm] {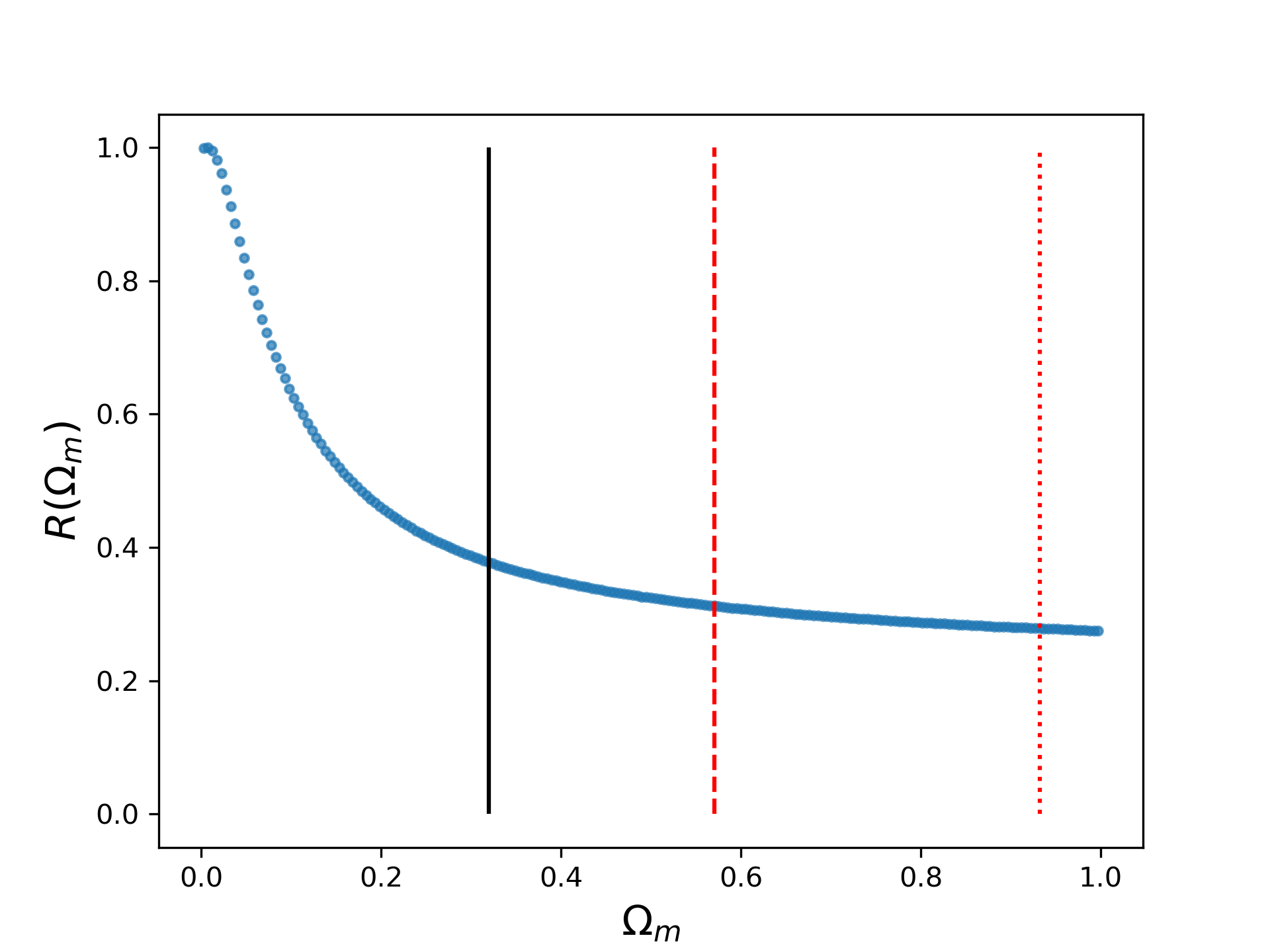}
\caption{{$R(H_0)$ and $R(\Omega_m)$ distributions for 7 CC data points with $z > 1.2$ ($z_{\textrm{min}} = 1.2$ from Table \ref{tab:LCDM_CC}). The dashed and dotted lines denote $1 \sigma$ and $2 \sigma$ confidence intervals. The black line denotes the best fit parameters for the full sample.}}
\label{fig:R_zmin12}
\end{figure}

\section{A tension with Planck}
\label{sec:tension}
A $2 \sigma$ ($p = 0.021$) tension with Planck has been reported in OHD through best fits and mock simulations in \cite{Colgain:2022rxy}. In particular, it was noted that a combination of 7 CC and BAO data points above $z = 1.45$ resulted in a $(H_0, \Omega_m) = (37.8, 1)$ best fit, where in line with analysis here, an $\Omega_m \in [0, 1]$ uniform prior was assumed. Based on mock simulations, the probability of such a best fit configuration arising by chance in mocks assuming input parameters consistent with Planck was estimated to be $p = 0.021$ \cite{Colgain:2022rxy}. A similar best fit appears in the last entry of Table \ref{tab:LCDM_CC} and Table \ref{tab:LCDM_CC_PD}, but there is no tension with Planck within the errors, even with our PD analysis, because CC data is inherently of poorer quality than BAO data. One further difference between the analysis is that \cite{Colgain:2022rxy} imposes a Gaussian Planck prior $\Omega_m h^2 = 0.1430 \pm 0.0011$ \cite{Planck:2018vyg} \footnote{This prior essentially prevents high redshift CC data from tracking a non-evolving $H(z)$.} to fix the high redshift behaviour of $H(z)$, whereas our analysis here so far has not introduced a prior. 

\begin{figure}[htb]
   \centering
\includegraphics[width=80mm]{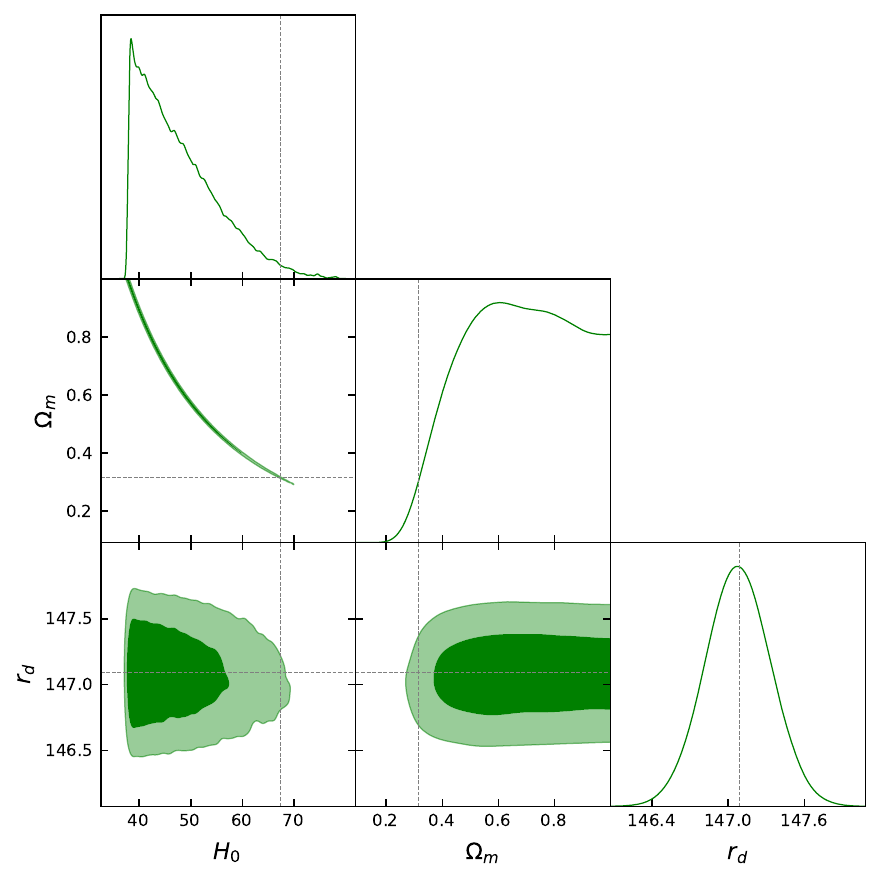}
\caption{MCMC posteriors for high redshift CC and BAO data subject to Planck priors on $\Omega_m h^2$ and $r_d$. Dashed lines mark the Planck central values $(H_0, \Omega_m, r_d) = (67.36, 0.315, 147.09).$}
\label{fig:CC_BAO_MCMC} 
\end{figure}

Nevertheless, armed with a new PD methodology, we are in a position to revisit the earlier result and see if we can recover the $2 \sigma$ tension with Planck. Since \cite{Colgain:2022rxy} made use of older BAO data, here we replace QSO and Lyman-$\alpha$ BAO with the latest eBOSS results \cite{Hou:2020rse, Neveux:2020voa, duMasdesBourboux:2020pck
}. Moreover, we work directly with the $D_{H}/r_d$ constraints and do not invert them. This entails assuming a value for the radius of the sound horizon, which we take to be the Planck value, $r_d = 147.09 \pm 0.26$ Mpc \cite{Planck:2018vyg}. In addition, we reinstate the prior $\Omega_m h^2 = 0.1430 \pm 0.0011$, so that the only difference with \cite{Colgain:2022rxy} is simply to update OHD BAO to the latest constraints. We stress that the priors we introduce are consistent with the Planck cosmology, so \textit{they cannot be driving any disagreement}. Moreover, the $\Omega_m h^2$ prior restricts one to a curve in the $(H_0, \Omega_m)$-plane, but it cannot dictate where one is on the curve, this is done by the remaining 3 CC and 3 BAO data points.  

We again marginalise over the free parameters $(H_0, \Omega_m, r_d)$ with MCMC. In Fig.~\ref{fig:CC_BAO_MCMC} we present the posteriors. While $r_d$ is Gaussian and peaked on our Planck prior, as expected, the $\Omega_m$ posterior is peaked at $\Omega_m \sim 0.6$ and the fact that the fall off in the distribution is gradual beyond the peak leads to a pile up of configurations in the top left corner of the $(H_0, \Omega_m)$-plane. This fall off continues beyond $\Omega_m = 1$ and if the prior is relaxed, the $H_0$ peak shifts to smaller values. So, once again all the hallmarks of projection effects are present. That being said, given the sharp fall off in the $\Omega_m$ distribution to smaller $\Omega_m$ values, some discrepancy appears to be evident with the Planck values (dashed lines). We remind the reader again that the $H_0$ posterior is unimpacted by our prior $H_0 \in [0, 200]$. However, we note that the 2D posterior in the $(H_0, \Omega_m)$-plane is largely unconstrained, so one suspects that projection effects are present in the $H_0$ posterior. What distinguishes the corner plot in Fig. \ref{fig:CC_BAO_MCMC} from the corner plots in Fig.\ref{fig:CCsplit1} and Fig. \ref{fig:CCzmin12} as genuine is that a peak in the $H_0$ posterior at \textit{smaller} values coincides with a peak in the $\Omega_m$ posterior at \textit{larger} values, thereby recovering the expected $H_0-\Omega_m$ anti-correlation in the $\Lambda$CDM model. In Fig. \ref{fig:CCsplit1}, where the focus is on the secondary phantom peak, and Fig. \ref{fig:CCzmin12} this anti-correlation is not evident in the $H_0$ and $\Omega_m$ posteriors. 

\begin{figure}[htb]
   \centering
\includegraphics[width=76.7mm] {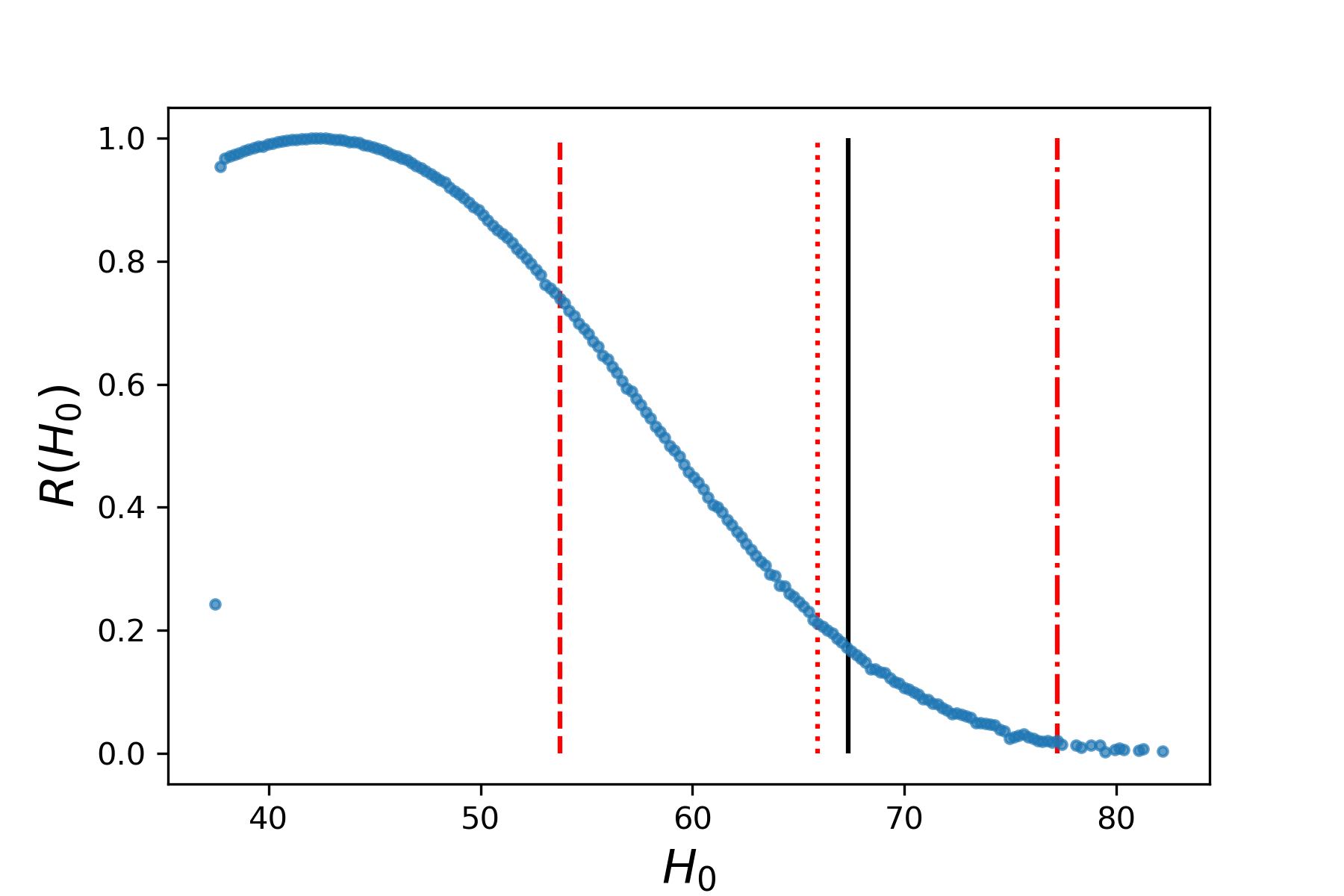}  \includegraphics[width=76.7mm] {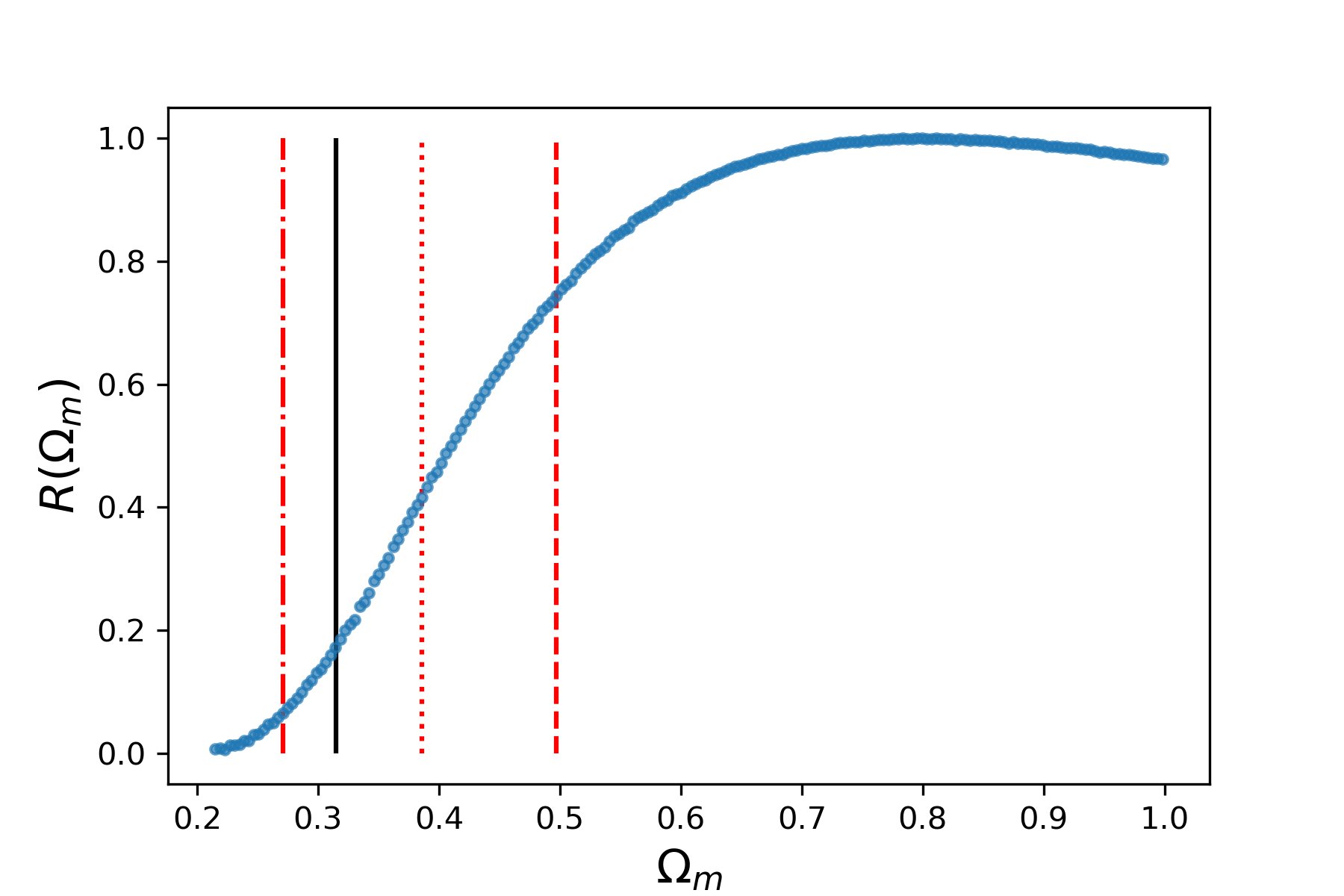}
\caption{$R(H_0)$ and $R(\Omega_m)$ distributions for CC and BAO OHD with redshift $z>1.45$ subject to Planck priors on $\Omega_m h^2$ and $r_d$. The dashed, dotted and dashed-dotted lines denote $68\%$, $95\%$ and $99.7\%$ confidence intervals and the solid black line is the Planck best fit value.}
\label{fig:CC_BAO}
\end{figure}

We now run the MCMC chain through our PD methodology. From Fig.~\ref{fig:CC_BAO}, we can see that the $R(H_0)$ and $R(\Omega_m)$ distributions prefer smaller values of $H_0$ and larger values of $\Omega_m$. The peak of the distributions occurs at $H_0 = 42.40$ km/s/Mpc and $\Omega_m = 0.795$.  The lone dot in the $R(H_0)$ distribution at low values of $H_0$ tells us that the distribution falls off sharply below $H_0 = 40$ km/s/Mpc. Note, since we employed generous uniform priors $H_0 \in [0, 200]$, the priors are not impacting the $R(H_0)$ distribution, so it is expected that the distribution falls off to zero on both sides. In contrast, the $R(\Omega_m)$ distribution is one-sided and fails to fall off in the direction of larger values within the uniform priors $\Omega_m \in [0, 1]$. The tension with Planck falls between {$95\%$ ($2 \sigma$) and $99.7\%$ ($3 \sigma$) confidence intervals}. By integrating the PDF as far as the black lines corresponding to the Planck values in Fig.~\ref{fig:CC_BAO}, we estimate that the Planck $H_0$ is located at the $96\%$ ($2.1 \sigma$) confidence level, while the Planck $\Omega_m$ value is at the $98.7\%$ ($2.5 \sigma$) confidence level. {We remind the reader that some care is required with $\Omega_m$ inferences because the PD is curtailed by the bounds. The same is true for MCMC posteriors with priors.}

The main take-away from this section is that OHD data comprising CC and BAO data points beyond $z=1.45$ is inconsistent with the Planck cosmology at in excess of $2 \sigma$. We have employed Planck priors to arrive at this result, but these priors cannot drive the disagreement. Moreover, independent analysis based on least squares fitting and mock simulations presented in \cite{Colgain:2022rxy} also points to a $2 \sigma$ tension, albeit with less up-to-date high redshift BAO data. In summary, different methodologies agree on a $2 \sigma$ discrepancy with Planck, which is robust to interchanging older and newer BAO data. 

\section{Concluding remarks}
\label{sec:discussion}
A $\chi^2$ likelihood is a metric or measure of how well a model fits data. The point in model parameter space that fits the data the best possesses the lowest $\chi^2$. Once one has identified this point, the problem remains to establish $68\%$, $95\%$, etc, confidence intervals in parameter space. In cosmology and astrophysics, MCMC is the prevailing technique for estimating credible intervals. Its great advantage is that it allows one to i) globally sample the parameter space and ii) arrive at posteriors that serve as an estimate of the errors even with non-Gaussian distributions. In contrast, if one minimises the $\chi^2$ by gradient descent, there is always a risk that one ends up in a local minimum, i. e. the global minimum is missed, while error estimation through Fisher matrix essentially assumes any distribution is Gaussian. The appeal of MCMC marginalisation is that it is widely applicable. However, the point of this paper is that limitations exist, even in the simplest model. 

Indeed, what happens when the MCMC posterior {is impacted by a degeneracy and} no longer tracks points in parameter space that fit the data the best? Traditionally, volume effects are seen as the preserve of higher-dimensional models, e. g. \cite{Herold:2021ksg, Gomez-Valent:2022hkb, Meiers:2023gft}, but projection effects also occur in the minimal $\Lambda$CDM model when one fits the model to data binned by redshift in the late Universe \cite{Colgain:2022tql}. As explained in \cite{Colgain:2022tql}, this ``projection effect'' is driven by OHD, $H(z_i)$, and angular diameter or luminosity distance data, $D_{A}(z_i)$ or $D_{L}(z_i)$, respectively only constraining the combinations $\Omega_m h^2$ and $ (1-\Omega_m) h^2$ well, with high redshift data $z_i \gg 0$. In practice, this restricts MCMC configurations to constant $\Omega_m h^2$ and constant $(1-\Omega_m) h^2$ curves in the $(H_0, \Omega_m)$ plane, and as the curves stretch due to DE or matter being less well constrained in high redshift bins, projection effects lead to shifts in the peaks of MCMC posteriors and the emergence of non-Gaussian tails \cite{Colgain:2022tql}. We stress that one sees the same effect in PDFs of best fit $(H_0, \Omega_m)$ parameters in a large number of mock data realisations \cite{Colgain:2022tql}, so the problem is more general than MCMC marginalisation; there is an inherent bias in the $\Lambda$CDM model when one fits it to redshift binned $H(z)$ \textit{or} $D_{A}(z)$ \textit{or} $D_{L}(z)$ data. {Succinctly put, the $\Lambda$CDM model confronted to redshift-binned OHD must transition from a 2D model to an effective 1D model. This impacts both Bayesian and frequentist methods, but the effect is more pronounced in MCMC marginalisation.} Within MCMC marginalisation, one sees this effect in the errors, the shift in peaks in 1D posteriors, but also in the drift of the parameters corresponding to the $\chi^2$ minimum outside of the $68\%$ credible intervals. Highlighting this (expected) feature in MCMC marginalisation using OHD is the opening salvo (result) of this paper.     

Why should one care? This is evidently only a problem if one bins data and confronts the $\Lambda$CDM model. First, note that some data sets are inherently binned. For example, effective redshifts are assigned to CC and BAO analysed in a given redshift bin, while each strongly lensed system constitutes its own bin. Working with binned data is unavoidable. Secondly, $\Lambda$CDM tensions point to a problem with the $\Lambda$CDM model once the tensions become widespread and persistent. As explained in \cite{Krishnan:2020vaf}, if the minimal $\Lambda$CDM model is too simple, one expects redshift evolution of $\Lambda$CDM fitting parameters as it is confronted to redshift binned data. Hints of these trends are now evident in $H_0$ \cite{Wong:2019kwg, Millon:2019slk, Dainotti:2021pqg, Colgain:2022nlb, Colgain:2022rxy, Malekjani:2023dky, Hu:2022kes, Jia:2022ycc, Krishnan:2020obg, Dainotti:2022bzg}, $\Omega_m$ \cite{Risaliti:2015zla, Risaliti:2018reu, Lusso:2020pdb, Yang:2019vgk, Khadka:2020vlh, Khadka:2020tlm, Khadka:2021xcc, Pourojaghi:2022zrh, Colgain:2022nlb, Colgain:2022rxy, Malekjani:2023dky, Pasten:2023rpc, Sakr:2023hrl} and $S_8$/$\sigma_8$ \cite{Esposito:2022plo, Adil:2023jtu, ACT:2023dou, ACT:2023kun} (also \cite{Miyatake:2021qjr, Alonso:2023guh}) across a host of different observables. This evolution in fitting parameters is an expected hallmark of model breakdown, which must happen at some redshift if systematics are not universally at play. 

The main problem with redshift dependent $\Lambda$CDM fitting parameters \footnote{There is a separate interpretation problem as the cosmology literature works with parameters ``defined today''. In more mathematical language, as explained in the introduction, this is simply the statement that one solves an ordinary differential equation (ODE), namely the Friedmann equation or equivalent, by specifying an integration constant, e.g. $H_0 = H(z=0)$ or $\rho_m(z=0)=\rho_{m0}=H_0^2\Omega_{m}$. However, this is a mathematical statement and it still needs to be confirmed observationally that $H_0$ or $\rho_{m0}$ are \textit{bona fide} constants from the perspective of observation. This cannot be \textit{a priori} assumed, because it is mathematical prediction of the model. If the model is correct, confronting the $H_0$ and $\Omega_m$ fitting parameters to data at different redshifts will return consistent constant values of the parameters. This is an important observational consistency test of the $\Lambda$CDM model. See \cite{Krishnan:2020vaf} for further discussion.} is one needs to assign a statistical significance to any trend. At a purely practical level, this entails constructing bins centered on different redshifts and identifying discrepancies in $\Lambda$CDM fitting parameters between bins, \textit{ideally in the same observable}, so that the potential systematics are under greatest control. As demonstrated both mathematically and observationally with the CC data in section \ref{sec:MCMC_bias}, MCMC marginalisation leads to biased inferences from projection effects when one bins the data. In this paper we have resorted to profile distributions \cite{Gomez-Valent:2022hkb} to overcome this bias and have applied the technique to a setting where {projected 1D $\Lambda$CDM posteriors} are expected to be non-Gaussian for the reasons outlined above and in section \ref{sec:MCMC_bias}. This new technique, provides a complementary perspective that confirms the least square fits of observed and mock data presented in \cite{Colgain:2022nlb, Colgain:2022rxy, Malekjani:2023dky}, where evidence for redshift evolution in $H_0$ and $\Omega_m$ was presented. Regardless of the methodology, the objective is to drill down on the prevailing \textit{assumption} that $\Lambda$CDM fitting parameters are constants. \textit{In the era of tensions in cosmology, nothing can be assumed, especially noting that the tensions are in essence showing an example of evolution of these fitting parameters with redshift and hence the model breakdown.}

More concretely, in this paper with both mock simulations and profile distributions we have shown that high redshift CC data has a preference for a non-evolving $H(z)$ over Planck-$\Lambda$CDM at approximately $\sim 2 \sigma$. This trend, which constitutes the second result of the paper, is unquestionable, as it is visible in the data. Note, we have not propagated systematic uncertainties, so the significance will be less when these are properly propagate. Nevertheless, low and high redshift CC data currently have a preference for different $\Lambda$CDM fitting parameters. Interestingly, as we have shown in section \ref{sec:MCMC_puzzle}, resorting exclusively to MCMC marginalisation one cannot arrive at this conclusion because 2D posteriors are unconstrained, and as a result, uninformative. It is important that observables at all redshifts return similar $H_0$ values because if the CC program is claiming an 8\% constraint on the Hubble constant, $H_0 = 66.7 \pm 5.5$ km/s/Mpc \cite{Moresco:2023zys}, it is imperative that \textit{all subsets of the data are consistent with this result}. If they are not, then we are staring at either systematics or model breakdown. Admittedly, demanding self-consistency of subsets of a data set confronted to a model is a high bar, but it is important that data sets result in overlapping constraints on $\Lambda$CDM parameters, otherwise this makes cosmological inferences moot. Note, the $\Lambda$CDM model is largely only well tested in the DE dominated regime $z \lesssim 1$ and at very high redshifts $z \sim 1100$, which leaves a wide expanse of redshifts to be explored in order to confirm or refute the model. Given the existing $\Lambda$CDM tensions \cite{Perivolaropoulos:2021jda, Abdalla:2022yfr}, and the hints of evolution in $H_0$, $\Omega_m$ and $S_8$ across assorted probes in the late Universe $z \lesssim 5$, it would be surprising if all discrepancies could be explained away by systematics.\footnote{We are open to the possibility, we just consider it a bad bet at the moment. The odds can of course change as observations improve.}

As an aside, it is intriguing that CC data has a preference for larger $H_0$ best fit values and smaller $\Omega_m$ best fit values beyond $z_{\textrm{min}} = 0.7$, as this is traditionally the transition redshift between decelerated and accelerated expansion. 
Moreover, at higher redshifts $z \sim 2.3$, there is not only a longstanding anomaly in Lyman-$\alpha$ BAO \cite{duMasdesBourboux:2020pck}, but QSOs also show a preference for a lower luminosity distance, $D_{L}(z)$, relative to Planck-$\Lambda$CDM \cite{Risaliti:2015zla, Risaliti:2018reu}. Translated into $\Lambda$CDM parameters, this corresponds to conversely larger $\Omega_m$ values, e. g.  \cite{Yang:2019vgk, Khadka:2020vlh, Khadka:2020tlm, Khadka:2021xcc, Pourojaghi:2022zrh}, and consequently smaller $H_0$ values. Thus, the emerging probes CC and QSOs  \cite{Moresco:2022phi} do not appear to be in sync on high redshift $\Lambda$CDM inferences. Nevertheless, neither may be inconsistent with the anomaly in Lyman-$\alpha$ BAO. Relative to Planck-$\Lambda$CDM, Lyman-$\alpha$ BAO prefers \textit{smaller} values of $D_{M}(z) := c \int_{0}^z 1/H(z^{\prime}) \, \textrm{d} z$ and \textit{smaller} values of $H(z)$ (larger values of $D_{H}(z) := c/H(z)$).\footnote{In this statement we assumed the Planck value $r_d \sim 147$ Mpc \cite{Planck:2018vyg} If we reinstate the radius of the sound horizon in these expressions, one recognises that changing the sound horizon, as advocated by early Universe resolutions to Hubble tension, cannot consistently address the Lyman-$\alpha$ BAO anomaly. In general, even for the Planck-$\Lambda$CDM sound horizon, one cannot get both a smaller $D_{M}(z)$ and smaller $H(z)$ from a strictly increasing function, such as the $\Lambda$CDM $H(z)$. As a result, deviations from the Planck-$\Lambda$CDM model that address this anomaly are expected to lead to wiggles in $H(z)$ \cite{Akarsu:2022lhx}, which are unsurprisingly seen in data reconstructions \cite{Zhao:2017cud, Wang:2018fng, Escamilla:2021uoj}. Finally, evolution in $H_0, \Omega_m$ discussed here cannot be explained or accommodated by early resolutions to Hubble tension relying on a change in the $r_d$ at very high $z$.}. If CC data prefer less evolution in $H(z)$ in the matter-dominated regime, then this is consistent with the preference for a smaller $H(z)$ from Lyman-$\alpha$ BAO. Furthermore, QSO data prefers smaller luminosity distances $D_{L}(z)$ relative to Planck, which are consistent with the smaller $D_{M}(z) \propto D_{L}(z)$ values preferred by Lyman-$\alpha$ BAO. Thus, even if CC and QSOs appear to be showing diverging behaviour in the cosmological fitting parameters $(H_0, \Omega_m)$, this may still turn out to be consistent with Lyman-$\alpha$ BAO. We await future DESI \cite{DESI:2023ytc} data releases to ascertain if the non-evolving $H(z)$ trend in high redshift CC data is physical or not. 

Finally, we come to our third and main result outlined in section \ref{sec:tension}. We have revisited a $\sim 2 \sigma$ tension between high redshift CC and BAO data reported in \cite{Colgain:2022rxy}, where the significance was estimated through mock simulations. Here, we have upgraded the BAO data to the latest constraints and again  recover a $>2 \sigma$ discrepancy in $(H_0, \Omega_m)$ with different methodology. This provides a consistency check that there is evolution in cosmological fitting parameters from OHD between low and high redshifts in the late Universe. Note, this evolution runs contrary to the non-evolving $H(z)$ seen in high redshift CC data because it assumes Planck has accurately constrained the high redshift behaviour of the Hubble parameter in (\ref{eq:lcdm}). Nevertheless, both with and without a Planck prior on $\Omega_m h^2$, evolution at $ \gtrsim 2 \sigma$ is evident in OHD data. It should be stressed that evolution is evident in PDFs of best fit $\Lambda$CDM parameters fitted to a large number of Planck-$\Lambda$CDM mocks \cite{Colgain:2022tql}, so evolution in fitting parameters from observed data can be expected. It is imperative to revisit the remaining observations in \cite{Colgain:2022rxy, Malekjani:2023dky} in order to confirm the significance of $\sim 2 \sigma$ hints of evolution found separately in Type Ia SNe and QSO data sets.

\section*{Acknowledgments}
We thank Adri\`a G\'omez-Valent and Leandros Perivolaropoulos for discussions and comments on the draft. We thank Gabriela Marques, Mike Hudson and Matteo Viel for related discussions on late Universe evolution in $S_8$. E\'OC thanks Yonsei University and Asia Pacific Center for Theoretical Physics for hospitality. 
This article/publication is based upon work from COST Action CA21136 – “Addressing observational tensions in cosmology with systematics and fundamental physics (CosmoVerse)”, supported by COST (European Cooperation in Science and Technology).  MMShJ thanks the support from ICTP associates office (under Senior Associate program) and ICTP HECAP section for hospitality. SP and MMShJ acknowledge SarAmadan grant No. ISEF/M/401332.

\appendix
\section{Fisher Matrix}
\label{sec:fisher}
Consider the $\chi^2$ (\ref{eq:chi2}). 
Defining 
\begin{equation}
    H_{\textrm{model}}(z) = H_0 \sqrt{1-\Omega_m + \Omega_m (1+z)^3}, 
\end{equation} and $Q_i$ as in \eqref{eq:Q}, we can now work out the derivatives
\begin{equation}
    \begin{split}
\partial_{H_0} Q_i &= -\sqrt{1-\Omega_m + \Omega_m (1+z_i)^3}, \\  \partial_{\Omega_m} Q_i &= - \frac{1}{2} H_0 (z_i^3 + 3 z_i^2 + 3 z_i)/\sqrt{1-\Omega_m + \Omega_m (1+z_i)^3}, \\
\partial^2_{H_0} Q_i &= 0, \\
\partial_{H_0} \partial_{\Omega_m} Q_i &= - \frac{1}{2} (z_i^3 + 3 z_i^2 + 3 z_i)/\sqrt{1-\Omega_m + \Omega_m (1+z_i)^3}, \\
\partial^2_{\Omega_m} Q_i =& \frac{1}{4} H_0 (z_i^3 + 3 z_i^2 + 3 z_i)^2/(1-\Omega_m + \Omega_m (1+z_i)^3)^{\frac{3}{2}}.      
    \end{split}
\end{equation}
We can then define the Fisher matrix 
\be
F_{ij} = \frac{1}{2} \frac{\partial^2 \chi^2(H_0, \Omega_m)}{\partial p_i \partial p_j}
\ee
where $p_i \in \{ H_0, \Omega_m \}$. Note that the Fisher matrix is evaluated on the best fit parameters. The result is a $2 \times 2$ matrix, which one inverts and the estimated errors are the square root of the diagonal entries.

\end{document}